\documentclass[letterpaper,12pt]{article}
\usepackage{tabularx} 
\usepackage{amsmath}  
\usepackage{graphicx} 
\usepackage[margin=1in,letterpaper]{geometry} 
\usepackage{natbib}
\setcitestyle{authoryear,open={(},close={)}}
\usepackage[final]{hyperref} 
\usepackage{subfigure}
\hypersetup{
	colorlinks=true,       
	linkcolor=blue,        
	citecolor=blue,        
	filecolor=magenta,     
	urlcolor=blue         
}

\begin{document}

\title{Numerical investigation of mixed-phase turbulence induced by a plunging jet}
\author{Rong Li, Zixuan Yang, Wei Zhang}
\date{\today}
\maketitle

\begin{abstract}
       In nature and engineering applications, plunging jet acts as a key process 
       causing interface breaking and generating mixed-phase turbulence. 
       In this paper, high-resolution
       numerical simulation of water jet plunging into a quiescent pool was perform to investigate the statistical property of the mixed-phase turbulence, with special focus on the closure problem of the Reynolds-averaged equation. 
       We conducted phase-resolved simulations, with the air--water interface 
       captured using a coupled level-set and volume-of-fluid method. 
       Nine cases were performed to analyse the effects of the Froude number and Reynolds number. 
       The simulation results showed that the turbulent statistics are insensitive to the Reynolds number under investigation, while the Froude number influences the flow properties significantly. 
       To investigate the closure problem of the mean momentum equation, the turbulent kinetic energy (TKE) and turbulent mass flux (TMF) and their transport equations were further investigated. 
       It was discovered that the balance relationship of the TKE budget terms remained similar to many single-phase turbulent flows. 
       The TMF is an additional unclosed term in the mixed-phase turbulence. 
       Our simulation results showed that the production term of its transport equation was highly correlated to the TKE.
       Based on this finding, a closure model of its production term was further proposed.

\end{abstract}


\section{Introduction}
\label{sec_Introduction}

Mixed-phase turbulence is a common phenomenon in nature and 
engineering applications. 
Difference from two-phase turbulent flow without surface breaking 
~\citep{brocchiniDynamicsStrongTurbulence2001}, 
mixed-phase turbulence is accompanied by violent surface deformation 
and surface breakup, leading to complex mass and momentum transfer 
across interface. 


Water jet plunging is a typical process that induces mixed-phase turbulence  
~\citep{kigerAirEntrainmentMechanismsPlunging2012,delacroixExperimentalStudyBubble2016}. 
For some advanced ships, plunging jet adds another source of air entrainment 
in the transom region and beyond~\citep{hsiaoNumericalExperimentalStudy2013}.
Plunging liquid jets were early investigated in chemical engineering applications,
such as waste water treatment~\citep{inproceedings}, 
mixing and reacting of liquids and gases~\citep{MCKEOGH19811161}. 
The research in this area has focused on the mechanism of air entrainment 
and the characteristics of the resulting bubble flows. 
%
It was found that the impact velocity of the liquid jet
plays a dominant role in the air entrainment inception conditions. 
It was reported that the entrainment velocity is a key 
criterion of air entrainment~\citep{lorenceauAirEntrainmentViscous2004}. 
When the critical condition was reached, a minimum amount of energy was 
available in the flow to do work against surface tension and/or 
the potential energy of gravity to entrap air.
Laboratory experiments were conducted to analyse the critical entrainment 
velocity for both high-viscosity liquids
~\citep{BIN19933585,josephTwodimensionalCuspedInterfaces1991,
	jeongFreesurfaceCuspsAssociated1992,eggersAirEntrainmentFreeSurface2001,
	lorenceauFractureViscousLiquid2003,lorenceauAirEntrainmentViscous2004} 
and low-viscosity liquids~\citep{BIN19933585,SENE19882615,linDROPSPRAYFORMATION1998,
	elhammoumiMeasurementsAirEntrainment2002,chirichellaIncipientAirEntrainment2002}.
%
~\citet{clanetDepthPenetrationBubbles1997} proposed a model to predict 
the penetration depth of the air bubbles entrained by a water jet impacting 
into a flat water pool.
This model showed that the depth is determined by the initial jet momentum and 
the bubble terminal velocity as a function of its size.
Comprehensive reviews of the related research can be found in \citet{BIN19933585} 
and \citet{kigerAirEntrainmentMechanismsPlunging2012}.

The aforementioned studies focused on the vertical plunging jet, 
and recently, the horizontal and shallow-angle plunging jets 
have also been investigated. 
%
\citet{deshpandeComputationalExperimentalCharacterization2012} 
studied a shallow-angle plunging jet and revealed a periodic pattern of 
air entrainment, which does not happen when the impinging angle is steep. 
After that, \citet{deshpandeDistinguishingFeaturesShallow2013} 
corroborated that the periodicity scaled linearly with 
the Froude number through numerical simulations. 
They also studied both computationally and analytically the underlying 
causes responsible for large cavity formation at shallow angles. 
They found a strong stagnation pressure region for the shallow impacts, 
which serves to deflect the entire incoming jet flow radially outwards, 
producing a large cavity and subsequently creating splashing events. 
~\citet{hsiaoNumericalExperimentalStudy2013} 
studied a stationary and moving horizontal jet plunging into a quiescent water pool. 
Their numerical and experimental results showed that the frequency of air entrainment 
depended on the jet diameter and relative velocity with respect to the free surface.
%

In addition to the plunging jet, mixed-phase turbulence with air entrainment also 
occurs in hydraulic jump, wave breaking, near-surface shear flow and wake behind a column. 
\citet{CHACHEREAU2011896} 
conducted experimental studies on air entrainment in hydraulic jumps 
and observed that air entrainment takes place as the Froude number exceeds a critical value. 
They also found that the volume of air entrainment increases with an increasing Froude number.
\citet{garrettConnectionBubbleSize2000}, 
\citet{deaneScaleDependenceBubble2002}, 
\citet{wang_yang_stern_2016} 
and \citet{deike_melville_popinet_2016} 
studied the bubbles induced by breaking waves 
and observed that the bubble size spectrum is proportional to $ r^{-10/3} $, 
where $r$ is the effective radius of the bubbles. 
\citet{garrettConnectionBubbleSize2000} 
proposed a cascade scenario to understand the power law.
\citet{yuNumericalInvestigationShearflow2019} 
performed DNS of a canonical three-dimensional two-phase viscous turbulent flow 
with the turbulent kinetic energy (TKE) supplied by an underlying near-surface shear flow. 
They investigated the dependence of air entrainment and bubble size on 
the Froude number and Weber number.
They proposed a heuristic model that qualitatively matched and explained 
the salient evolution behaviour of the bubble size spectrum.
\citet{hendricksonWakeThreedimensionalDry2019} 
performed implicit large eddy simulations (iLES) with a conservative volume-of-fluid 
interface-capturing method to investigate the mixed-phase turbulent wake behind 
a dry transom stern of a surface ship. They conducted detailed analyse on the 
air entrainment including entrainment rate and bubble size spectrum. 

From the above reviews, it is understood that 
the mechanism and physical process of air entrainment and bubble generation 
have been studied extensively in the mixed-phase turbulence. 
However, the studies on the statistical characteristics and transport 
mechanics of the turbulence were limited in the literature.
As the most common phenomenon that induces two-phase turbulence in nature, 
breaking waves modulate the transfer of mass, momentum, and energy between 
the ocean and atmosphere. 
\citet{deikeMassTransferOcean2022} summarized the recent researches on breaking waves. 
Based on canonical experiments~\citep{rappLaboratoryMeasurementsDeepwater,EnergyDissipationbyBreakingWaves,
	melvilleVelocityFieldBreaking2002,bannerWaveBreakingOnset2007,
	drazenInertialScalingDissipation2008,tianEnergyDissipationTwodimensional2010}, 
simulations with direct numerical simulations (DNS)~\citep{chenTwodimensionalNavierStokes1999,iafratiNumericalStudyEffects2009,
	deikeCapillaryEffectsWave2015,deikeAirEntrainmentBubble2016,
	wangHighfidelitySimulationsBubble2016,
	yangDirectNumericalSimulation2018,
	chanTurbulentBubbleBreakup2021a,mostertHighresolutionDirectSimulation2022} 
and large eddy simulations (LES)~\citep{lubinNumericalSimulationsThreedimensional2015,
	derakhtiBreakingonsetEnergyMomentum2016}, 
the dynamics of wave breaking were investigated. 
However, as a statistically unsteady process, it is difficult to analyse 
the statistical properties of turbulence induced by breaking wave. 
To date, the research of turbulent statistics corresponding to breaking 
wave is limited to its impact on the wind turbulence~\citep{yangDirectNumericalSimulation2018}, 
and the study on the mixed-phase region is limited. 
In terms of statistically stationary mixed-phase turbulence, 
\citet{deshpandeComputationalExperimentalCharacterization2012} investigated 
the mean velocity, mean volume of fluid and the TKE for a plunging jet. 
\citet{yuNumericalInvestigationShearflow2019} investigated the characteristics of 
free-surface turbulence and elucidated the qualitatively distinct characteristics 
of strong versus weak free-surface turbulence. 
Strong anisotropy was found in weak free-surface flow while mixed-phase turbulence 
became almost isotropic in strong free-surface turbulence. 
When it comes to statistical averaging of variable density fluid motion equations, 
an additional unclosed term appears, that is, the turbulent mass flux 
(TMF) $\overline{\rho u_i^\prime}$~\citep{taulbeeModelingTurbulentCompressible1991}. 
Modelling of TMF was earlier investigated in compressible 
flow~\citep{1979pmtf.vkif.....J,grassoSupersonicFlowComputations1989,1990aiaa.meetW....N}. 
Comprehensive reviews can be found in \citet{chassaingModelingVariableDensity2001}. 
Recently, \citet{hendricksonWakeThreedimensionalDry2019a}
studied the incompressible highly variable density turbulence 
in the wake behind a three-dimensional dry transom stern. 
They developed an explicit algebraic closure model for the TMF. 

The objective of present study is to investigate the statistical characteristics of 
the mixed-phase turbulence with air entrainment caused by a plunging jet. 
Specifically, we aim to study in detail the basic statistics reflecting turbulent characteristics, 
such as TKE, TMF and their transport equations. 
The Froude number effect is examined through different cases. 
The closure problem of TMF is also discussed. 
The remainder of this paper is organized as follows. 
In section~\ref{sec_Details_of_numerical_simulations}, the numerical method and 
physical setup of present simulation of the plunging jet are described. Then, 
the results are presented and discussed in section~\ref{sec_Results_and_discussion}, 
followed by the conclusions in section~\ref{sec_Conclusion}.

\section{Details of numerical simulations}
\label{sec_Details_of_numerical_simulations}

\subsection{Numerical method}
\label{subsec_Numerical_method}

\noindent
We perform high-resolution large eddy simulation (LES) by solving the three-dimensional two-phase
incompressible Navier--Stokes equations to study a water jet plunging into a quiescent pool. 
The coupled level-set (LS) and volume of fluid (VOF) method is used to capture the air--water 
interface on a Cartesian grid. The three-dimensional incompressible Navier--Stokes equations
with varying density and viscosity are expressed as:

\begin{equation}\label{continuityEquation}
	\nabla \cdot \mathbf{u}=0 , 
\end{equation}
\begin{equation}\label{momentumEquation}
	\frac{\partial(\rho \mathbf{u})}{\partial t} + \nabla \cdot(\rho \mathbf{u} \mathbf{u}) 
	= -\nabla p + \nabla \cdot(2 \mu \mathbf{S}) + \rho \mathbf{g} + \mathbf{f}_s ,
\end{equation}

\noindent
where $ \rho $ and $ \mu $ are density and dynamic viscosity, respectively, 
$ \mathbf{u} = [u, v, w] $ is the velocity, $ p $ is the pressure, 
$ \mathbf{S} = (\nabla \mathbf{u} + \nabla \mathbf{u}^T) / 2 $ is the strain-rate tensor, 
$ \mathbf{g} = [0, -g, 0] $ is the gravitational acceleration. 
The last term in equation~(\ref{momentumEquation}) is the surface tension force, define as 

\begin{equation}\label{surfaceTensionForce}
	\mathbf{f}_s = \sigma \kappa \nabla H(\phi), 
\end{equation}

\noindent
where $ \sigma $ is the surface tension, 
$ \kappa = \nabla \cdot (\nabla \phi) $ at $ \phi = 0 $ is the interface curvature with 
$ \phi $ being the LS function, 
and $ H(\phi) $ is the Heaviside function, defined as

\begin{equation}\label{HeavisideFunction}
	H\left(\phi, \Delta_s\right) = 
	\begin{cases}
		0, & \phi \leq 0, \\ 
		1, & \phi > 0 .
	\end{cases}
\end{equation}

\noindent
In LES, the dynamic viscosity is expressed as 

\begin{equation}\label{dynamic_viscosity}
	\mu = \rho (\nu + \nu_t), 
\end{equation}

\noindent
where $\nu$ is the kinematic viscosity and $\nu_t$ the eddy viscosity. 
In present work, an eddy-viscosity subgrid-scale model for turbulent 
shear flow developed by~\cite{Vreman2004} is used to determine $\nu_t$. 
We note here that in low Reynolds number cases, it shows $\nu_t \ll \nu$, 
and as a result, $\nu_t$ is omitted in these cases  
following \cite{hendricksonWakeThreedimensionalDry2019}. 
Therefore, the case with low Reynolds numbers in the present work can be 
also regarded as high-resolution iLES. 

The interface between the two fluid phases is captured using the CLSVOF method. 
The following convection equations of the LS function $ \phi $ and 
VOF function $ \psi $ are solved:

\begin{equation}\label{lsEquation}
	\frac{\partial \phi}{\partial t}+\nabla \cdot(\phi \mathbf{u})=0, 
\end{equation}
\begin{equation}\label{vofEquation}
	\frac{\partial \psi}{\partial t}+\nabla \cdot(\psi \mathbf{u})=0. 
\end{equation}

\noindent
The LS function $ \phi $ is defined as the signed distance from 
either fluid phase to the interface; its sign is negative and positive 
for air and water, respectively.
The VOF function $ \psi $ is defined as the volume fraction of water 
in a grid cell ranging from 0 to 1. 
Let subscripts $ a $ and $ w $ denote air and water, respectively. 
The density and viscosity are determined using the LS function as follows:

\begin{equation}\label{density}
	\rho=\rho_a+\left(\rho_w-\rho_a\right) H(\phi), 
\end{equation}
\begin{equation}\label{viscosity}
	\mu=\mu_a+\left(\mu_w-\mu_a\right) H(\phi). 
\end{equation}

\noindent
As noted in the literature~\citep{rudmanVolumetrackingMethodIncompressible1998,
	arrufatMassmomentumConsistentVolumeofFluid2018,
	nangiaRobustIncompressibleNavierStokes2019,yangRobustSolverIncompressible2021}, 
if the density is determined solely 
using equation~(\ref{density}), the simulation is unstable 
for two-fluid flows with a high density contrast. 
An effective approach for improving the numerical stability is to 
calculate the mass and momentum fluxes using a consistent scheme 
and evolve the density by solving the following convection equation:

\begin{equation}\label{densityEquation}
	\frac{\partial \rho}{\partial t}+\nabla \cdot(\rho \mathbf{u})=0. 
\end{equation}

\noindent
In the present study, equation~(\ref{density}) is used to determine the density 
at the beginning of each time step, while equation~(\ref{densityEquation}) is 
evolved together with the momentum equation to provide the density within a time step. 
As such, the interface is accurately captured, and meanwhile, the simulation is stable.
More details about the numerical method can be found in 
~\citet{yangRobustSolverIncompressible2021}.

\subsection{Physical setup}
\label{subsec_Physical_setup}

\begin{figure}
	\centering
	\includegraphics[scale=0.5]{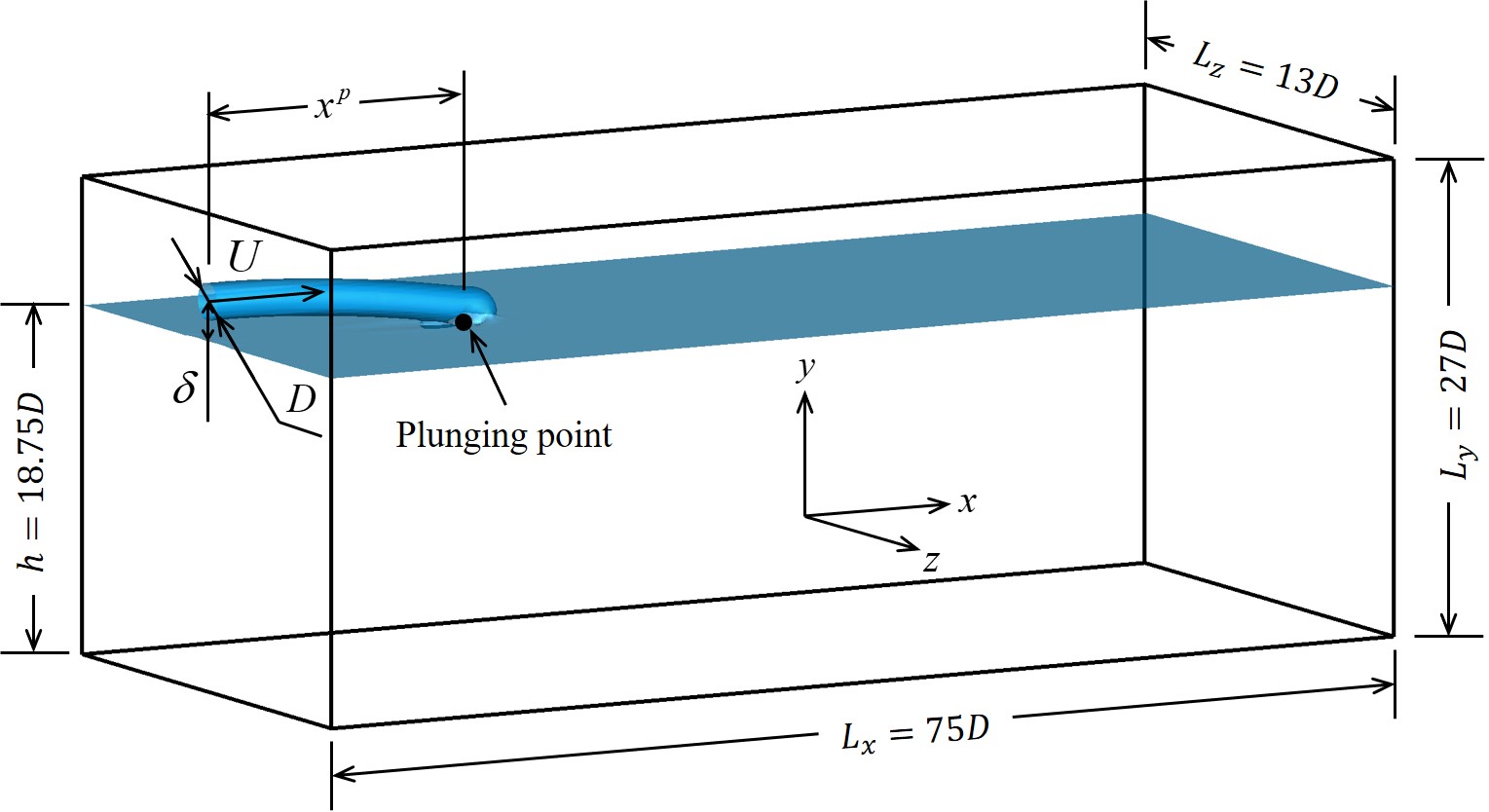} 
	\caption{\label{computation_domain} 
		Definition sketch of computational model.
	}
\end{figure}

\noindent
Figure~\ref{computation_domain} schematically shows the computational domain 
and definition of key parameters. 
As shown, a water jet spurts out horizontally and plunges into a quiescent water pool. 
Except for the water column and pool, other space of the computational 
domain is initially filled with air, some of which tends to be entrained 
into the water with the jet and then evolves into air cavities and bubbles 
beneath the free surface.
We use the jet diameter $ D $ as the characteristic length scale
and the horizontal outlet velocity of jet $ U $ as the characteristic velocity scale.
Hereafter, all variables are non-dimensionalized using $ U $ and $ D $ unless otherwise stated.
The computational domain size is $ L_x \times L_y \times L_z = 75.0 \times 27.0 \times 13.0 $,
where $ x$, $y$, $z $ represent streamwise, vertical and spanwise directions of the domain, respectively.
The water depth is $ h = 18.75 $. The distance between the centre 
of jet orifice and the free surface is $ \delta = 1.0 $. 
The boundary condition is no-penetration at the top, bottom, front and back of the domain, 
while an constant inlet velocity is applied in the jet orifice area of the left boundary 
and a zero gradient condition is specified at the right outlet boundary.
The whole computational domain is discretized using a uniform Cartesian grid 
and the grid resolution in three directions is $ \Delta x = \Delta y = \Delta z = 0.1 $.
In the present study, to consider the effect of the Reynolds number $ \text{Re} = {UD}/{\nu} $ 
and the Froude number $ \text{Fr} = \sqrt{{u^2}/{gD}} $, we conduct 9 cases. 
Key parameters are listed in table~\ref{table_parameters}. 
Cases 1, 2, 6 are performed to examine the effect of Reynolds number 
and cases 3-9 are conducted to investigate the Froude number effect.
The Reynolds number and Froude number for case 1 remain the same as the case 
considered in~\citet{deshpandeComputationalExperimentalCharacterization2012}
to facilitate validation.

\begin{table}
	\caption{\label{table_parameters} 
		Parameters in the simulations of plunging jet.
	}
	\begin{center}
		\begin{tabular}{*{10}{c}}
			\hline
			case & & $ \text{Re} $ & & $ \text{Fr} $ & & $ \text{We} $ \\
			\hline
			1 & & $1.6 \times 10^5$ & & 6.4 & & $ 8.89 \times 10^3 $ \\
			2 & & $1.6 \times 10^4$ & & 6.4 & & $ 8.89 \times 10^3 $ \\
			3 & & $1.6 \times 10^3$ & & 3.2 & & $ 8.89 \times 10^3 $ \\
			4 & & $1.6 \times 10^3$ & & 4.2 & & $ 8.89 \times 10^3 $ \\
			5 & & $1.6 \times 10^3$ & & 5.3 & & $ 8.89 \times 10^3 $ \\
			6 & & $1.6 \times 10^3$ & & 6.4 & & $ 8.89 \times 10^3 $ \\
			7 & & $1.6 \times 10^3$ & & 7.5 & & $ 8.89 \times 10^3 $ \\
			8 & & $1.6 \times 10^3$ & & 8.6 & & $ 8.89 \times 10^3 $ \\
			9 & & $1.6 \times 10^3$ & & 9.6 & & $ 8.89 \times 10^3 $ \\
			\hline
		\end{tabular}
	\end{center}
\end{table}


\section{Results and discussion}
\label{sec_Results_and_discussion}

\begin{figure}
	\centering
	\includegraphics[scale=0.15]{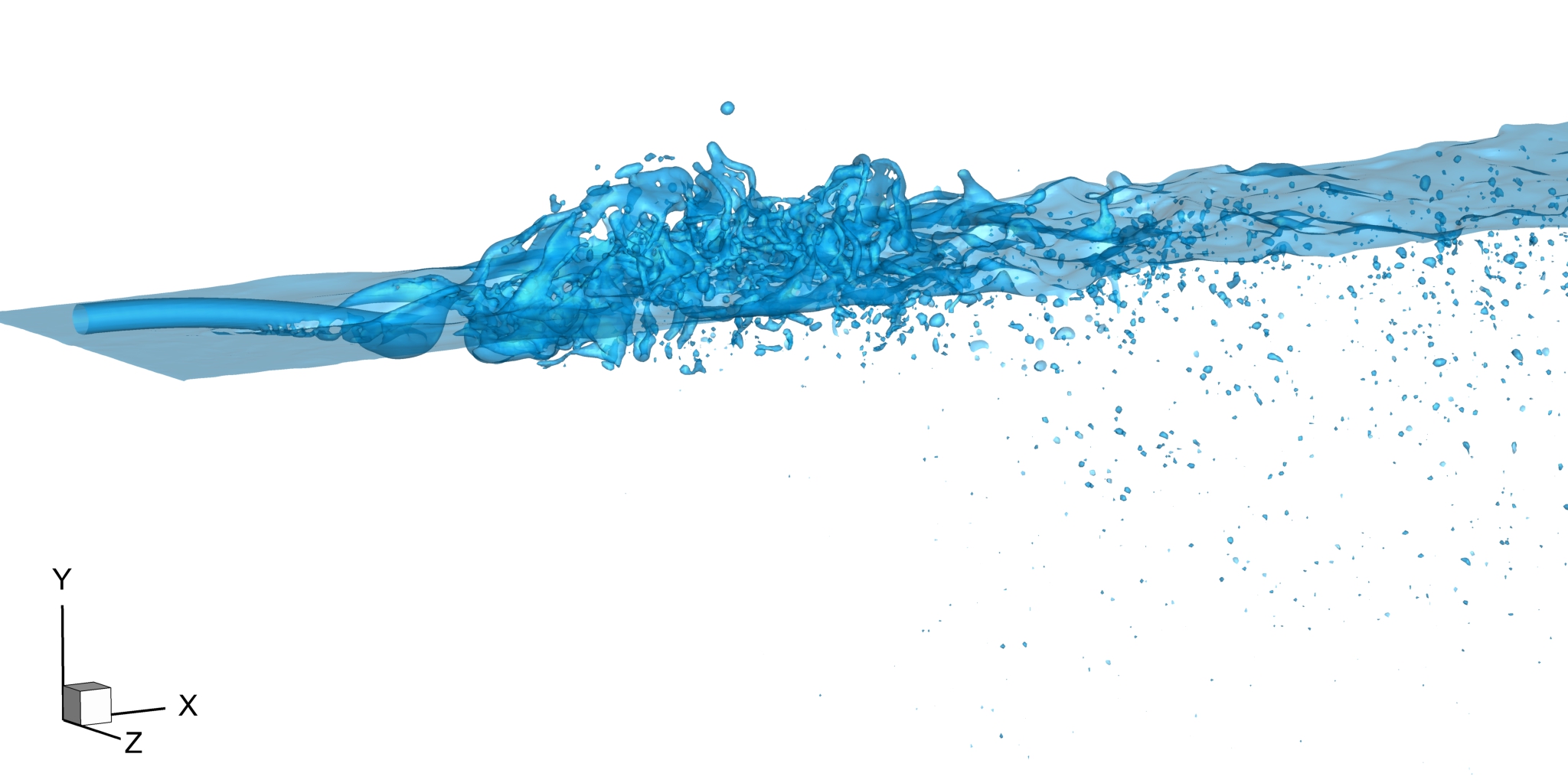} \hspace{+5mm}
	\caption{\label{computational_model} 
		Instantaneous air--water interface at a certain time under the statistical steady state 
		(iso-surface with $\phi = 0$ in case 6).
	}
\end{figure}

\subsection{Formation of mixed-phase region}
\label{subsec_Formation_of_mixed_phase_region}
\begin{figure}
	\centering
	\includegraphics[scale=0.35]{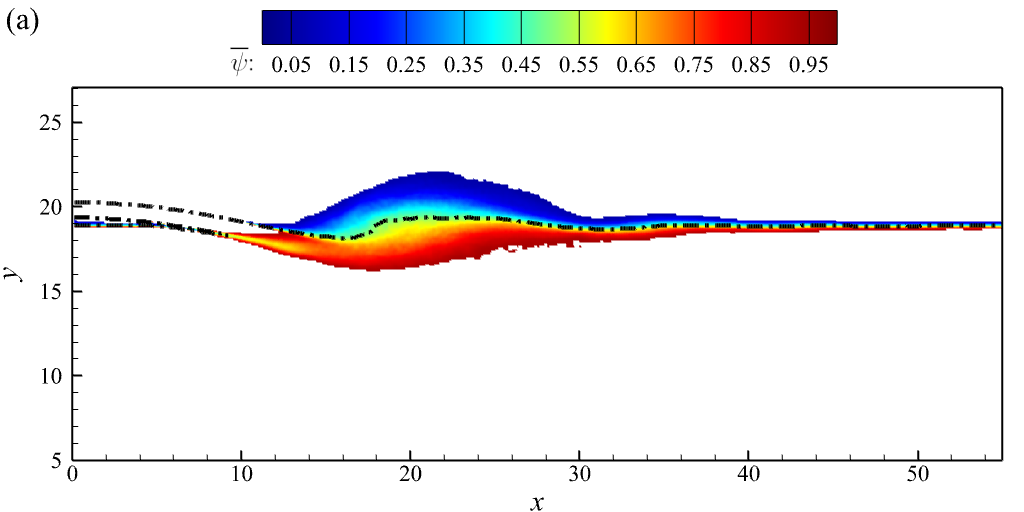} \hspace{+5mm}
	\includegraphics[scale=0.35]{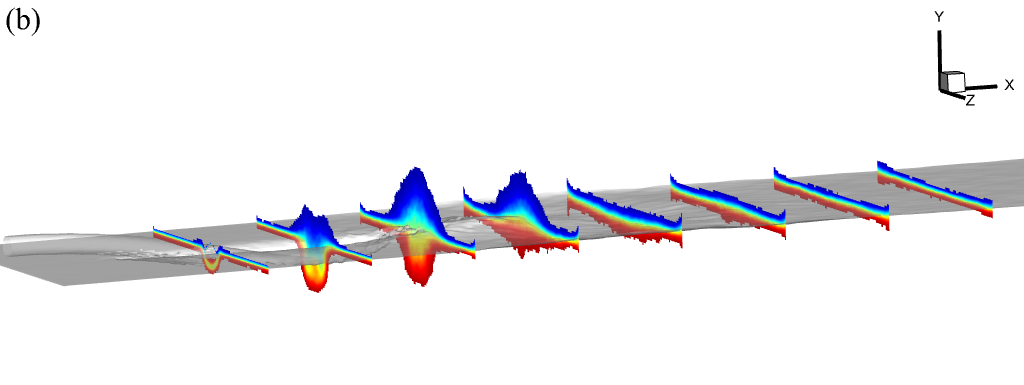}
	\caption{\label{FM_MPR} 
		Mixed-phase region in case 6:
		(a) Contour of mean volume of fluid $ 0.05 < \overline{\psi} < 0.95 $ 
		at the mid span.
		(b) Transverse cuts of mixed-phase region in different streamwise position. 
		Dash-dotted line in (a) and iso-surface in (b) represent the mean free surface 
		with $\overline{\psi} = 0.5$.
	}
\end{figure}

An instantaneous flow field under the statistically steady state is shown 
in figure~\ref{computational_model}, in which the interface between air and water 
is visualized using the iso-surface of $\phi = 0$. 
This figure illustrates entrained air pockets and bubbles under the surface, 
as well as water splash and droplets above the surface.
It can be seen that a large air pocket is entrained into the water near 
the plunging point, and it breaks up into small air pockets 
and bubbles in the downstream.
Above the free surface, water splash and droplets hit the downstream 
surface causing secondary plunging.
%
The plunging event results in violent breaking of the free surface and 
highly mixed air--water turbulent flow. 
To better understand the highly mixed air--water turbulence in the near-surface region, 
we follow~\citet{hendricksonWakeThreedimensionalDry2019a} to define a mixed-phase region
as the variable density region, 
where the mean volume of fluid $\overline{\psi}$ satisfies $ 0.05 \le \overline{\psi} \le 0.95 $. 
Here, the overline defines time averaging, 
which is performed over a time duration of $ T = 1200.0 $, 
after the turbulence is fully developed. 
The sampling rate is $\Delta T = 1.0$, which provides 1200 samples for time averaging.
Figure~\ref{FM_MPR} shows the mixed-phase region and mean free surface in case 6. 
It can be seen that as the streamwise coordinate $x$ increases, 
the size of the mixed-phase region increases and reaches a peak shortly after 
the jet plunging point and then decreases downstream. 
The mean free surface with $\overline{\psi} = 0.5$ is also shown in 
figure~\ref{FM_MPR}(a) using the dash-dotted line.
There exists a hollow of the mean free surface 
near the jet plunging point and a hump shortly downstream. 
They correspond to air entrainment and water splash-up, respectively. 

\subsection{The Reynolds number effect}
\label{subsec_The_Reynolds_number_effect}

The Reynolds number is an important parameter in turbulent flow. 
However, in many previous studies of mixed-phase 
turbulence~\citep{brocchiniDynamicsStrongTurbulence2001,
	deikeCapillaryEffectsWave2015,yuNumericalInvestigationShearflow2019}, 
it is found that the Reynolds number effect is less significant than 
the Froude number effect. 
To minimize the effect of LES modelling, 
it is a common treatment to reduce the Reynolds number. 
In this study, we test three Reynolds numbers to find 
its effects on turbulent statistics. 
Meanwhile, we compare these results with previous experimental 
and numerical studies to validate our simulations. 

\begin{figure}
	\centering
	\includegraphics[scale=0.6]{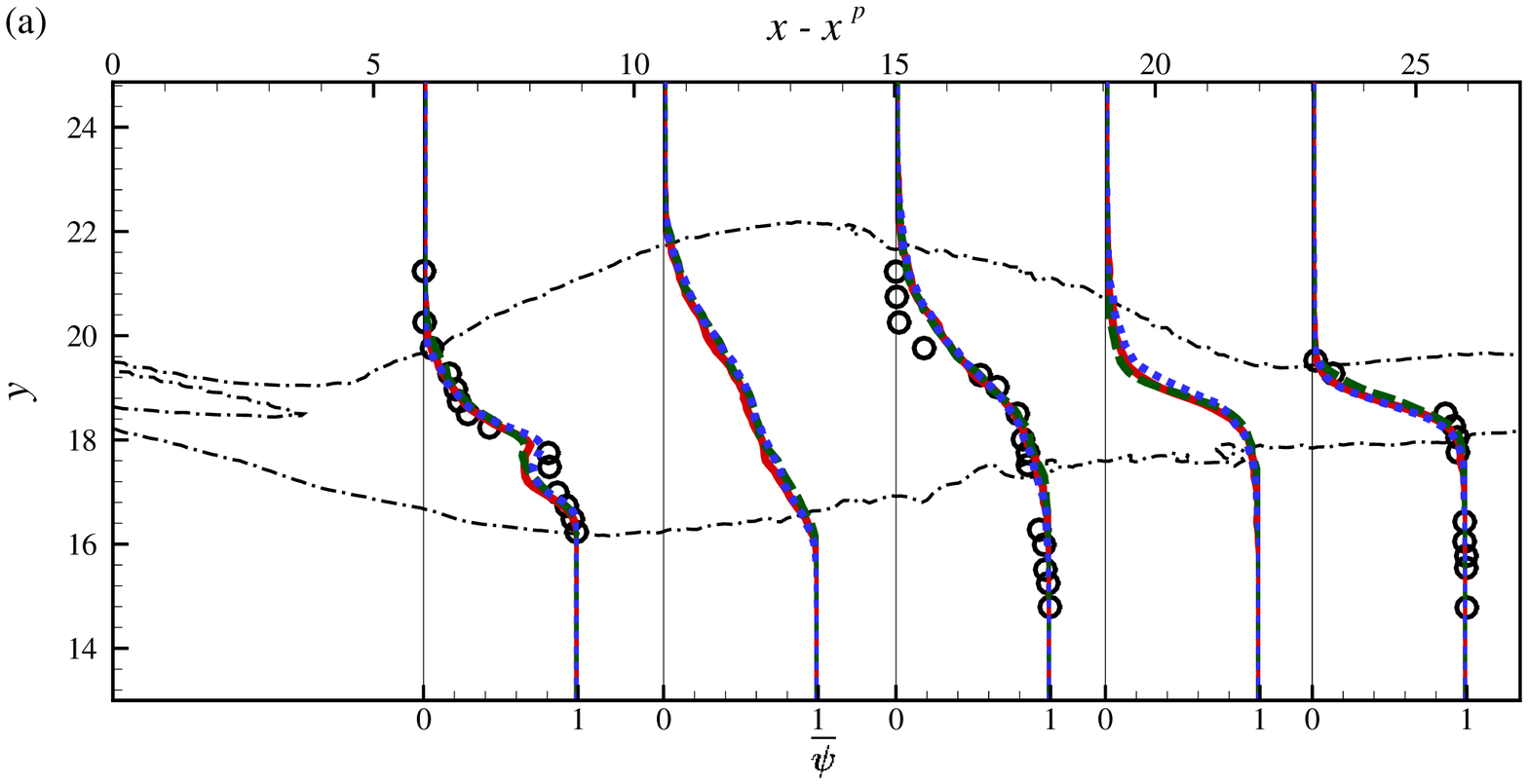} 
	\includegraphics[scale=0.6]{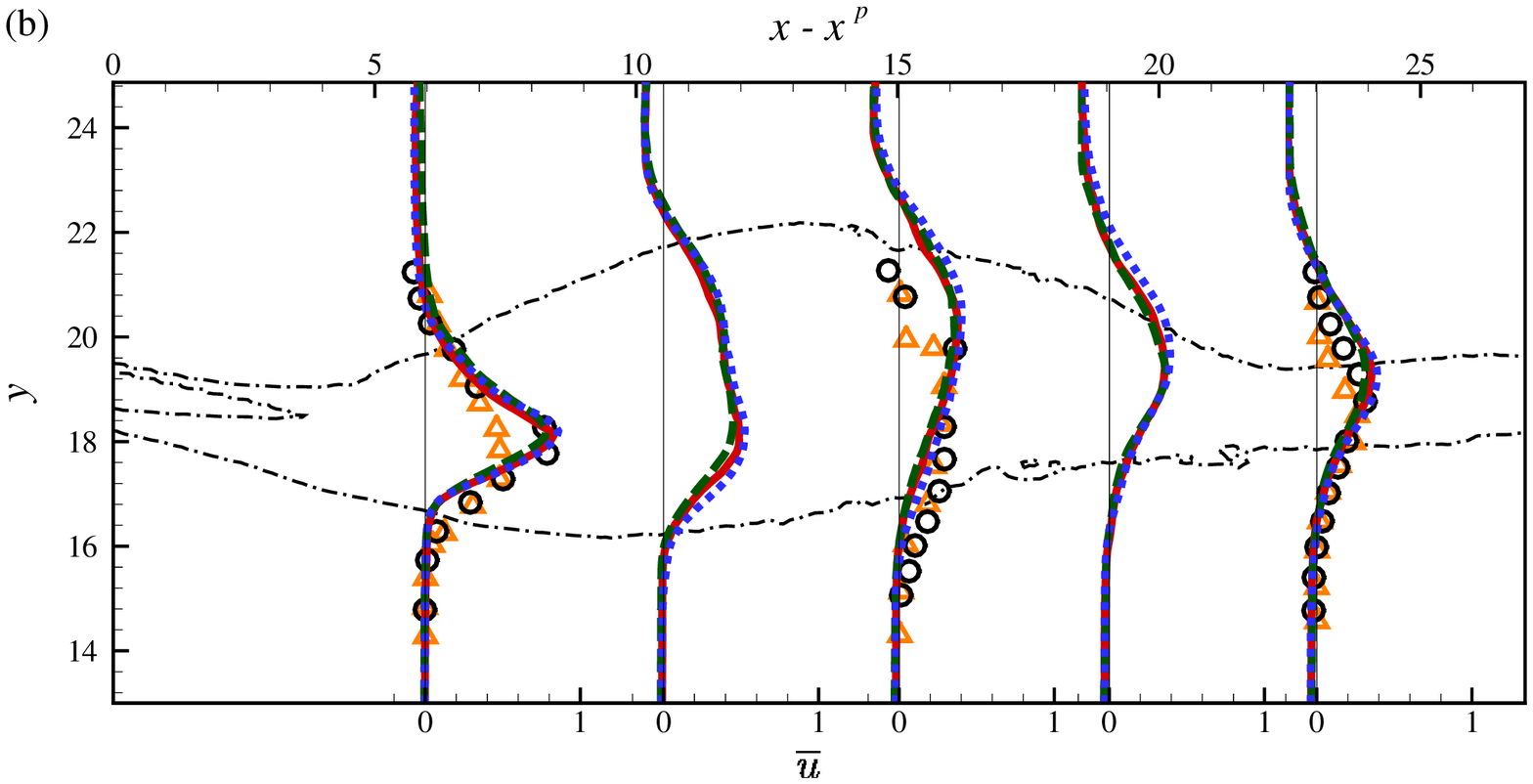} 
	\caption{\label{FM_UM_y_different_x_5_Re} 
		Vertical profiles of (a) mean volume of fluid and (b) mean 
		streamwise velocity at the mid span and different streamwise 
		locations for different Reynolds numbers.  
		Solid: $ \text{Re} = 1.6 \times 10^5 $, 
		dashed: $ \text{Re} = 1.6 \times 10^4 $, 
		dotted: $ \text{Re} = 1.6 \times 10^3 $, 
		circles and triangles represent numerical and experimental data, 
		respectively, of same case with $\text{Re} = 1.6 \times 10^5$ in 
		~\citet{deshpandeComputationalExperimentalCharacterization2012}. 
		Dash-dotted lines represent the edge of the mixed-phase region 
		in case 6.
	}
\end{figure}

Figure~\ref{FM_UM_y_different_x_5_Re} shows the vertical profiles 
of the mean volume of fluid $\overline{\psi}$ and mean streamwise 
velocity $\overline{u}$ at the mid span and different streamwise 
locations for the three Reynolds numbers. 
As shown in figure~\ref{FM_UM_y_different_x_5_Re}(a), 
the mean volume of fluid $\overline{\psi}$ varies mainly inside 
the mixed-phase region. 
The results of $\overline{\psi}$ for different 
Reynolds numbers are close with each other. 
They also agree with the numerical results of 
\citet{deshpandeComputationalExperimentalCharacterization2012}.
Figure~\ref{FM_UM_y_different_x_5_Re}(b) shows that 
the results of the mean velocity for different 
Reynolds numbers are also close with each other, 
which indicates that variation in the Reynolds number 
(from $ 1.6 \times 10^3 $ to $ 1.6 \times 10^5 $) 
does not impose significant effects on the mean flow.
The results also agree with the experimental and numerical results 
of \citet{deshpandeComputationalExperimentalCharacterization2012}. 

\begin{figure}
	\centering
	\includegraphics[scale=0.4]{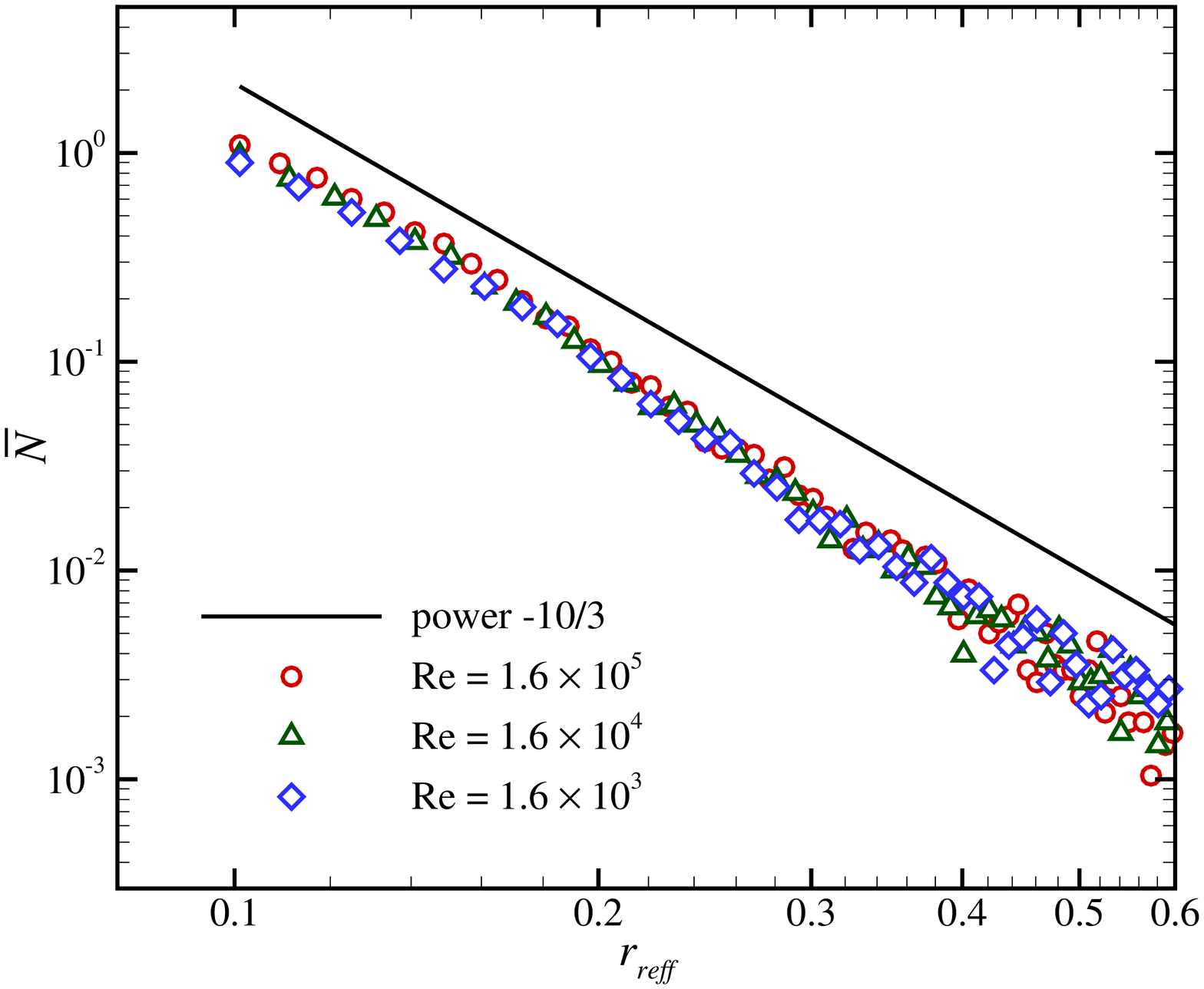} \hspace{+5mm}
	\caption{\label{N_R_different_Re} 
		Average bubble-size density spectra as a function of effective radius 
		in cases of different Reynolds number. ($\text{Fr} = 6.4$)
	}
\end{figure}

We also calculate the time-averaged bubble-size density spectra 
$ \overline{N}(r_{eff}) $, which is defined as 

\begin{equation}\label{def_of_N_R_eff}
	\overline{N}\left( r_{eff} \right) = 
	\frac{1}{T} \int_t^{t+T} 
	\frac{n\left(r_{eff}, t ; b\right)}{V \cdot b} \mathrm{~d} t, 
\end{equation}

\noindent
where $ n\left(r_{eff}, t ; b\right) $ is the number of bubbles, 
whose effective radii fall between $ r_{eff} $ and $ r_{eff} + b $ 
in a given fluid volume $ V $ at time $ t $. 
In the present work, $b=0.001$ is chosen. 
The fluid volume $V$ for bubble statistics is a cuboid in the 
computational domain of $ x \in [15, 45], y \in [0, h], z \in [0, L_z] $, 
which contains most of the air cavities and bubbles beneath the interface.
The effective spherical radius is defined as 

\begin{equation}\label{effective_radius_of_bubble}
	r_{eff} = \left[(3 / 4 \pi) v_e \right]^{1 / 3}, 
\end{equation}

\noindent
where $ v_e $ is the volume of an individual bubble. 
To determine the number and volume of each bubble, 
a connected component algorithm~\citep{sametEfficientComponentLabeling1988} 
is used to identify and label the entrained air cavities. 
Figure~\ref{N_R_different_Re} shows the results of 
the bubble-size density spectra $\overline{N}(r_{eff})$ 
for different Reynolds numbers. 
It can be seen that the results for different Reynolds 
numbers are close to each other, indicating that the Reynolds number 
effect on the bubble-size density spectra is negligible.
The solid line in figure~\ref{N_R_different_Re} represents 
the $ r^{-10/3} $ power law, which is satisfied in cases at 
different Reynolds numbers. 

The results of the mean velocity, volume of fluid, and 
the bubble-size density spectra for different Reynolds numbers 
(ranging from $1.6 \times 10^3$ to $1.6 \times 10^5$) 
indicate that the Reynolds number imposes limited impact on 
turbulence statistics. 
We have also examined the Reynolds number effects 
on other turbulent statistics, including TKE and TMF. 
It is found that the impact of the Reynolds number is 
less significant than the Froude number. 
Therefore, in the following context, 
we focus on the effect of the Froude number. 

\subsection{Scaling of cross-sectional area of mixed-phase region with Froude number}
\label{subsec_Scaling_of_cross_sectional_area_of_mixed_phase_region_with_Froude_number}

\begin{figure}
	\centering
	\includegraphics[scale=0.6]{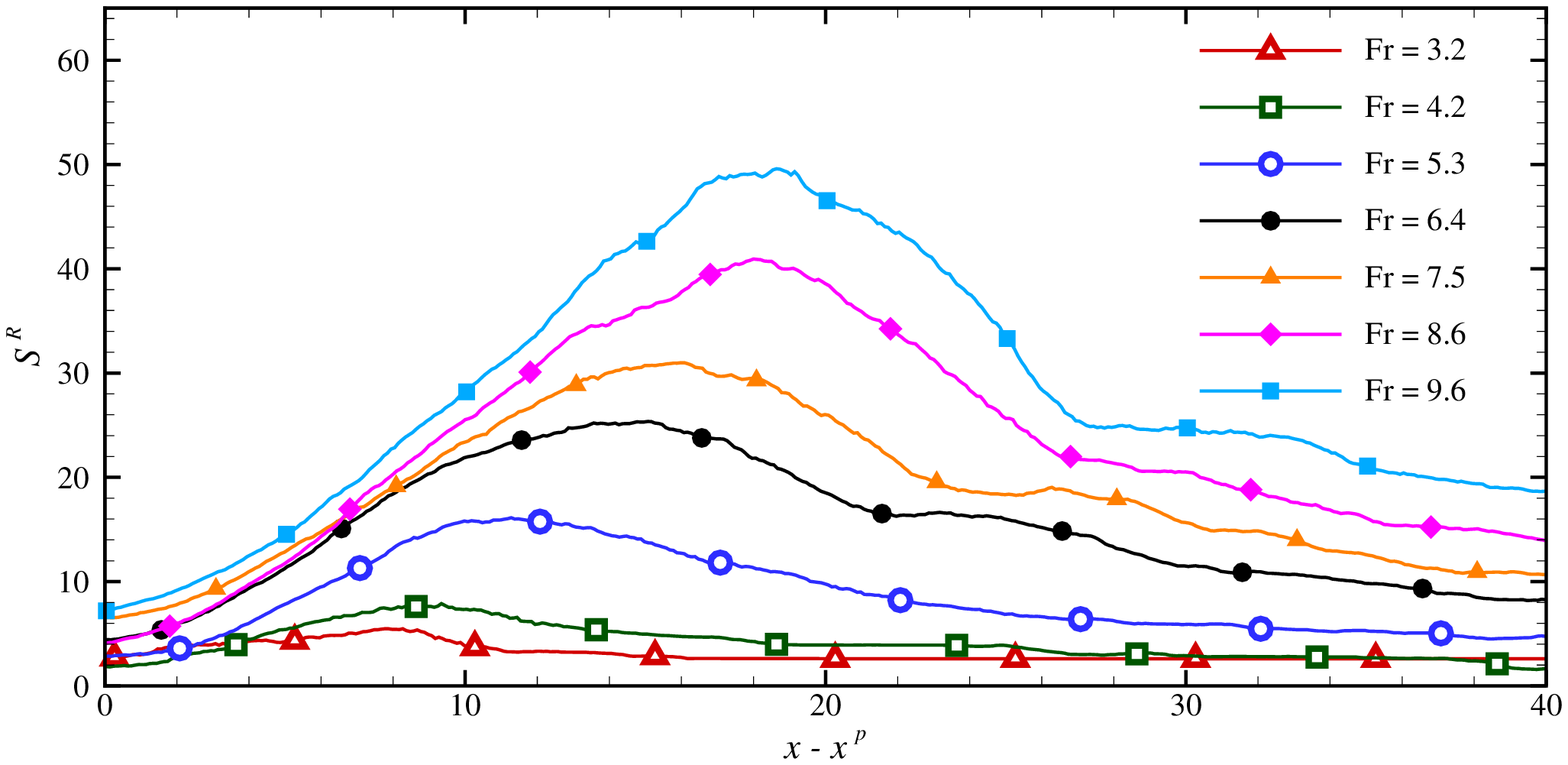}
	\caption{\label{SR_x-xp_different_Fr} 
		Profiles of the mixed-phase region area along the streamwise direction 
		in cases of different Froude number.
	}
\end{figure}
\begin{figure}
	\centering
	\includegraphics[scale=0.4]{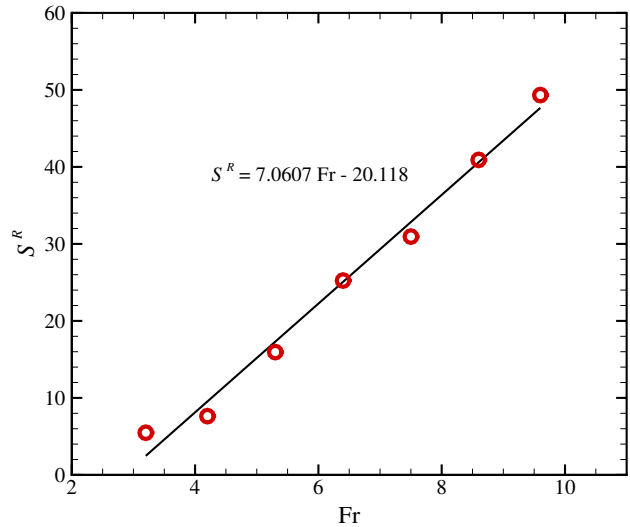}
	\caption{\label{SR_Fr} 
		Peak variables of the mixed-phase region area varying with Froude number and the fit curve.
	}
\end{figure}

We start analysing the Froude number effect from the size of the mixed-phase region. 
Figure~\ref{SR_x-xp_different_Fr} shows the the streamwise variation 
of the cross-sectional area $S^R(x)$ of the mixed-phase region. 
The jet plunging point $x^p$ is influenced by the Froude number, and we use $x - x^p$ 
as the independent variable to facilitate comparisons among different cases.
As shown in figure~\ref{SR_x-xp_different_Fr}, as the Froude number increases, 
the maximum cross-sectional area increases. 
Figure~\ref{SR_Fr} shows the variation of the maximum cross-sectional area of 
the mixed-phase region, $\text{max}(S^R)$, with respect to the Froude number. 
It is seen that $\text{max}(S^R)$ increases approximately in a linear law with the Froude number. 
The observations from figures~\ref{SR_x-xp_different_Fr} and \ref{SR_Fr} indicate 
that as the Froude number increases, the mixing of air and water is enhanced. 
Similar conclusion was drawn in previous studies of plunging 
jet~\citep{chirichellaIncipientAirEntrainment2002a,kigerAirEntrainmentMechanismsPlunging2012} 
and other mixed-phase turbulent flows, 
such as hydraulic jumps~\citep{maComprehensiveSubGridAir2011,CHACHEREAU2011896} 
and mixed-phase turbulence induced by shear near the interface~\citep{yuNumericalInvestigationShearflow2019}.

\subsection{Mean velocity}
\label{subsec_Mean_velocity}

\begin{figure}
	\centering
	\includegraphics[scale=0.35]{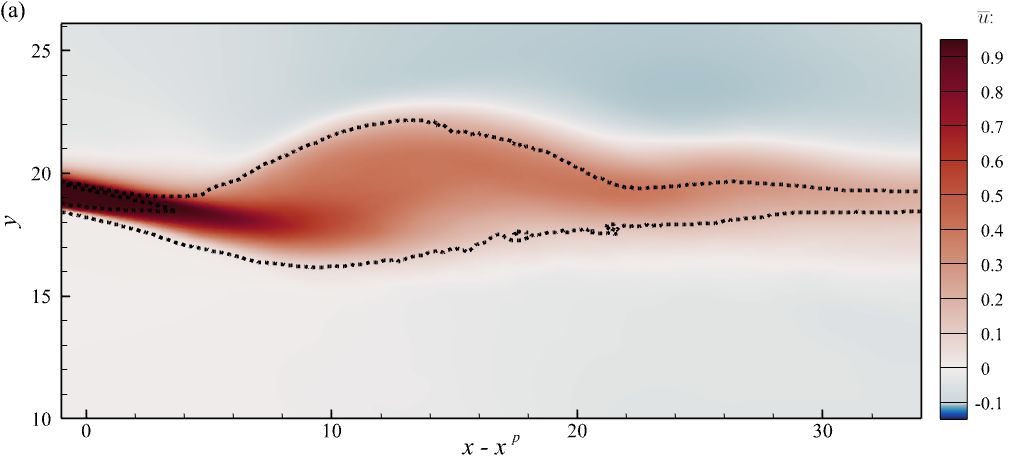}
	\includegraphics[scale=0.35]{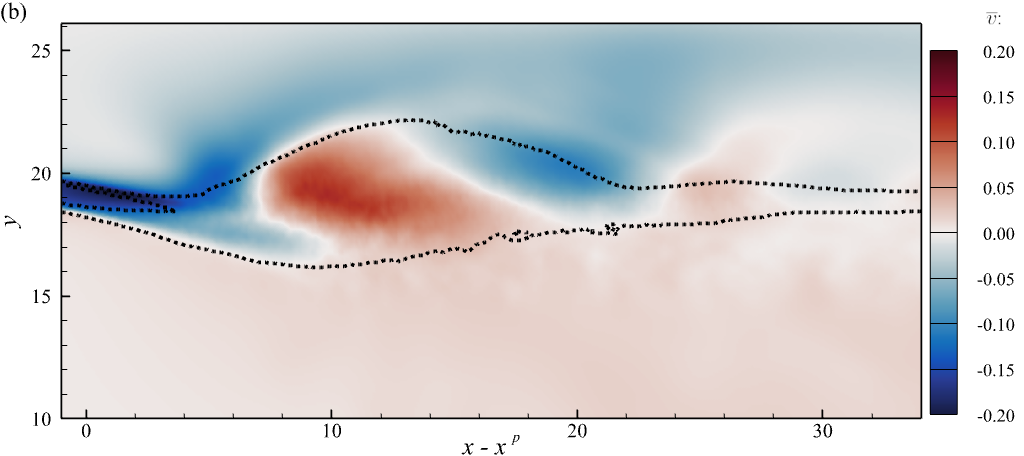}
	\caption{\label{UM_xy} 
		Contours of time-averaged velocities at the mid span: 
		(a), $\overline{u}$; (b), $\overline{v}$.
		Dotted lines represent the edge of the mixed-phase region.
		($ \text{Re} = 1600 $ and $ \text{Fr} = 6.4 $)
	}
\end{figure}

Figure~\ref{UM_xy} shows the contours of mean streamwise 
and vertical velocities at the mid span 
for case 6 at an intermediate Froude number $ \text{Fr} = 6.4 $. 
The dotted lines shows the upper and lower edge of the mixed-phase region. 
Figure~\ref{UM_xy}(a) shows that the jet plunging induces 
a mean streamwise velocity $\overline{u}$ near the free surface, 
and the large magnitude of $\overline{u}$ is collocated with the 
mixed-phase region. 
As shown in figure~\ref{UM_xy}(b), the trend of mean vertical 
velocity $\overline{v}$ varies along the streamwise direction 
in the mixed-phase region. Near the jet plunging point, 
the fluid around the jet moves downwards with it. 
Shortly downstream, the pool water moves upwards, and droplets are generated. 
Meanwhile, the air cavities under the surface moves upwards 
under the buoyancy. As a result, the vertical velocity is positive in this region. 
After the droplets reach the highest, 
they drop to the pool and form a secondary plunging, 
which causes another region with negative vertical velocity. 

Because the size of mixed-phase region varies in different Froude numbers, 
to facilitate comparison between results of different cases, 
we follow~\citet{hendricksonWakeThreedimensionalDry2019a} 
to define conditioned average in the mixed-phase region as:

\begin{equation}\label{Def_of_area_conditioned_average}
	\left\langle f^R \right\rangle _{y z}(x) = 
	\frac{1}{S^R(x)} \int_{y z} f^R(y, z ; x) \mathrm{d} S^R. 
\end{equation}

\noindent
Here, $f^R$ represents the variable $f$ inside the mixed-phase region. 
The integration denoted by $\left\langle \right\rangle _{y z}$ 
is performed over a cross-stream section, and $S^R(x)$ represents 
the area of the mixed-phase region in the cross-stream section.

\begin{figure} 
	\centering
	\includegraphics[scale=0.6]{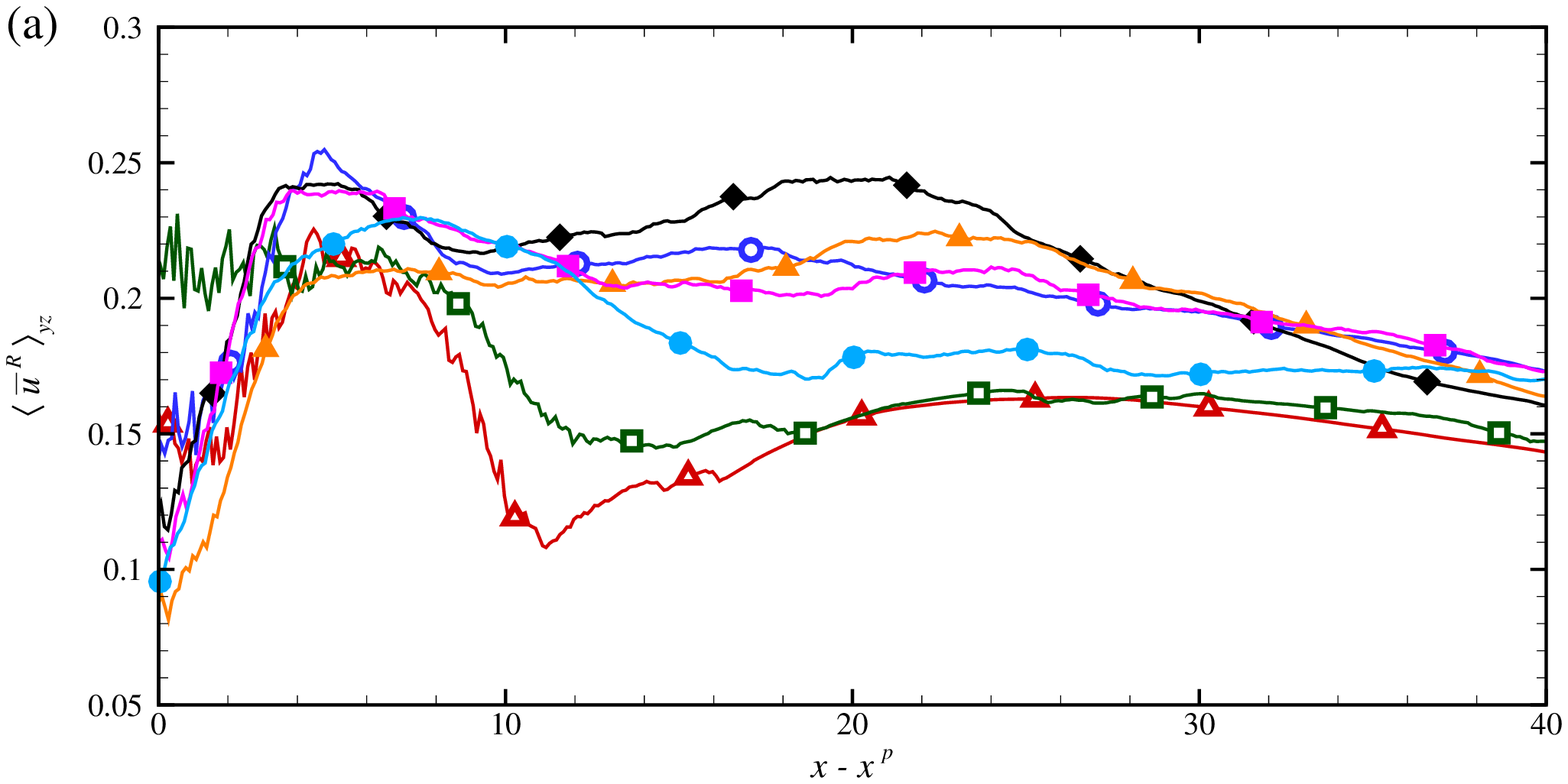}
	\includegraphics[scale=0.6]{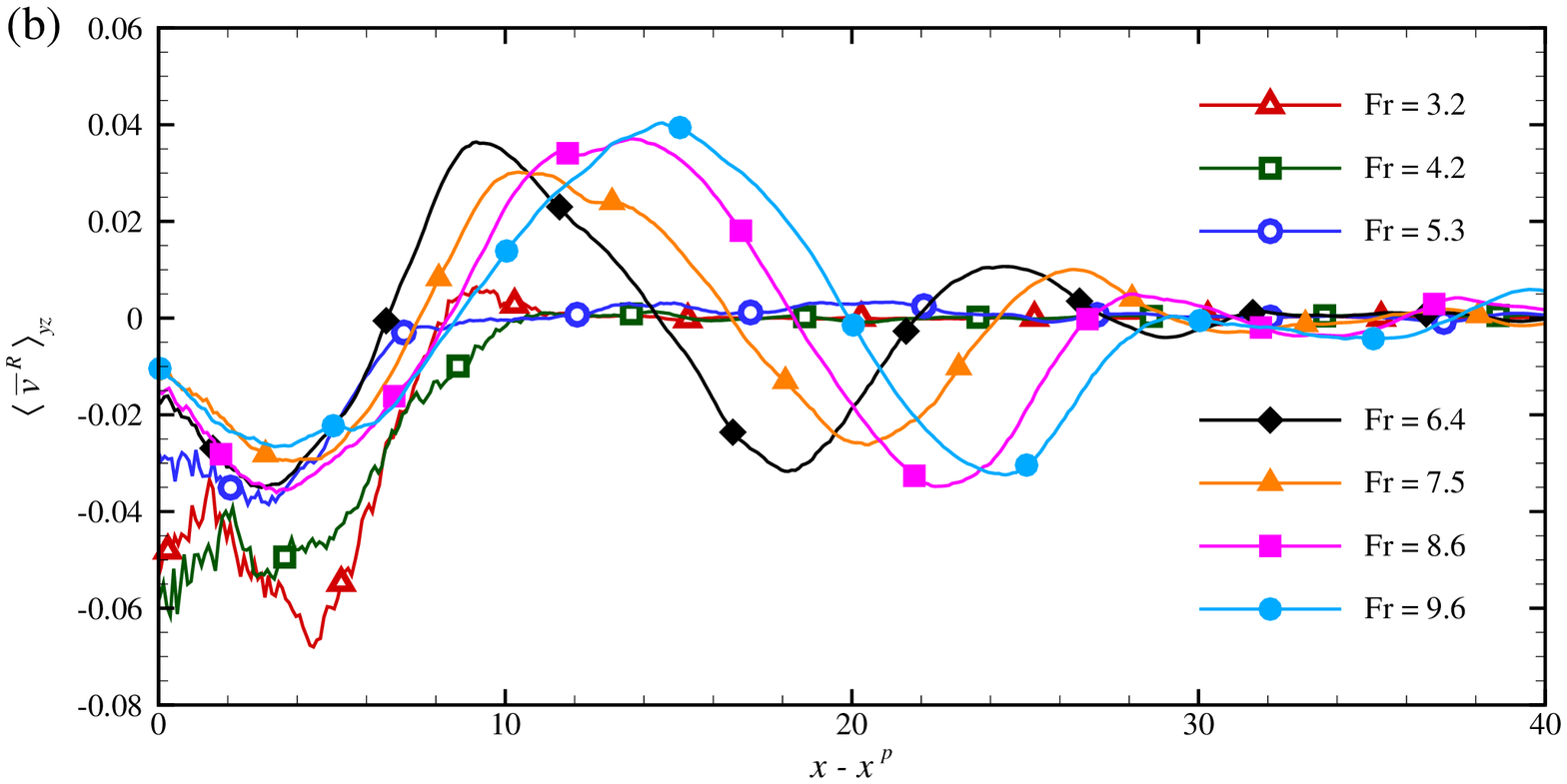}
	\caption{\label{UM_MPR_x-xp} 
		Mean velocities averaged in the mixed-phase region in different Froude numbers: 
		(a), $\left< \overline{u}^R \right>_{yz}$; (b), $\left< \overline{v}^R \right>_{yz}$.
	}
\end{figure}

Figure~\ref{UM_MPR_x-xp} compares the mean velocities averaged in the mixed-phase region
$\left< \overline{u}_i^R \right>_{yz}$ for different Froude numbers. 
We note here that the results in the flow region for $x - x^p = 0.0 \sim 4.0$ 
show some uncertainty, because the area of mixed-phase region $S^R(x)$ is small 
in this flow region. The reliability of statistics is questionable because of 
the small sampling number. Therefore, in the following content, 
we mainly focus on the rest flow region for $x - x^p > 4.0$, 
where the sampling number is sufficiently large to provide more reliable statistics. 
This does not influence our understanding of the statistical properties of the 
mixed-phase turbulence induced by the plunging jet, because active turbulence 
mainly occurs downstream, where air and water are sufficiently mixed. 

Figure~\ref{UM_MPR_x-xp}(a) shows that, as the value of velocity is 
non-dimensionalized by the jet horizontal velocity, 
$\left< \overline{u}^R \right>_{yz}$ shows little difference near the 
jet plunging region ($x - x^p = 4.0 \sim 8.0$) at different Froude numbers. 
This indicates that the mean flow is mainly induced by jet plunging, 
while turbulent motion does not impose significant influence on the 
mean flow in this region. Around $x-x^p=10.0$, where the splash is intense, 
the magnitude of $\left< \overline{u}^R \right>_{yz}$ reaches a valley. 
Downstream, $\left< \overline{u}^R \right>_{yz}$ shows a non-monotonic 
response to the increase of the Froude number. At lower Froude numbers 
($\text{Fr} \le 5.3$),$\left< \overline{u}^R \right>_{yz}$ 
increases with the Froude number. For $\text{Fr} \ge 6.4$, 
there exists intense vertical motion that expands the size of mixed-phase 
region, and as a result, the streamwise momentum is diffused and 
$\left< \overline{u}^R \right>_{yz}$ decreases as the Froude number increases. 

From figure~\ref{UM_MPR_x-xp}(b), it is observed that the magnitude 
of the first negative peak of $\left< \overline{v}^R \right>_{yz}$ 
around $x-x^p=4.0$ decreases as the Froude number increases because 
of the reduction of the gravitational potential energy of the injected 
water, which is proportional to $1/\text{Fr}$. 
Downstream of the plunging region, 
the vertical motion is weak at low Froude number for $\text{Fr} \le 5.3$, 
resulting in small magnitude of $\left< \overline{v}^R \right>_{yz}$. 
For $\text{Fr} \ge 6.4$, the water plunging induces a large amount 
of droplets, which induce the first positive peak of 
$\left< \overline{v}^R \right>_{yz}$. 
When the droplets reach the highest altitude, the vertical velocity 
becomes zero, and this is also the location where the size of 
mixed-phase region reaches the maximum. 
The secondary negative peak represents downward motion of droplets, 
which leads to the secondary plunging. 
Downstream the secondary plunging, there is still small magnitude 
of $\left< \overline{v}^R \right>_{yz}$ at higher Froude numbers. 
This indicates that the increasing Froude number results in more 
intense vertical motion of the surface. 

To perform statistical study of turbulent properties, 
an instantaneous variable $ f $ is decomposed as 
$ f(x, y, z; t) = \overline{f}(x, y, z) + f'(x, y, z; t) $, 
where $ f' $ is the fluctuation. 
The mean momentum equation of incompressible variable-density flow 
can be expressed as:

\begin{equation}\label{mean_momentum_equation}
	\frac{\partial (\overline{\rho} \overline{u}_i)}{\partial t} 
	= C_i + G^p_i + G_i + D^v_i + D^t_i + A_i. 
\end{equation}
Budget terms on the right-hand side of equation~(\ref{mean_momentum_equation}) 
include the convection term $ C_i $, 
pressure gradient term $G^p_i$, gravity term $G_i$, 
viscous diffusion term $D^v_i$,
Reynolds stress term $D^t_i$, 
and TMF term $A_i$. These terms are defined as
\begin{align}\label{mean_terms_def}
	&C_i = - \frac{\partial\left(\overline{\rho} \overline{u}_i \overline{u}_j\right)}{\partial x_j}, \\
	&G^p_i = - \frac{\partial \overline{P}}{\partial x_i}, \\
	&G_i = \frac{\overline{\rho}}{\text{Fr}^2} \delta_{i2}, \\
	&D^v_i = \frac{\partial \overline{\tau}_{i j}}{\partial x_j}, \\
	&D^t_i = - \frac{\partial \overline{\rho u_i^{\prime} u_j^{\prime}}}{\partial x_j}, \\
	\label{mean_terms_def_A}
	&A_i = - \frac{\partial \overline{\rho u_i^{\prime}}}{\partial t}
	- \frac{\partial}{\partial x_j}\left(\overline{\rho u_i^{\prime}} \overline{u}_j
	+ \overline{\rho u_j^{\prime}} \overline{u}_i\right),
\end{align}


\noindent
where $ \tau_{ij} = \mu \left( u_{i, j} + u_{j, i} \right) / \text{Re} $ is the 
viscous stress tensor. 
There are unclosed terms on the right hand side of equation~(\ref{mean_momentum_equation}), 
namely the Reynolds stress $\overline{\rho u_i^{\prime} u_j^{\prime}}$ 
and TMF $\overline{\rho u_i^{\prime}}$. 

\begin{figure}
	\centering
	\includegraphics[scale=0.6]{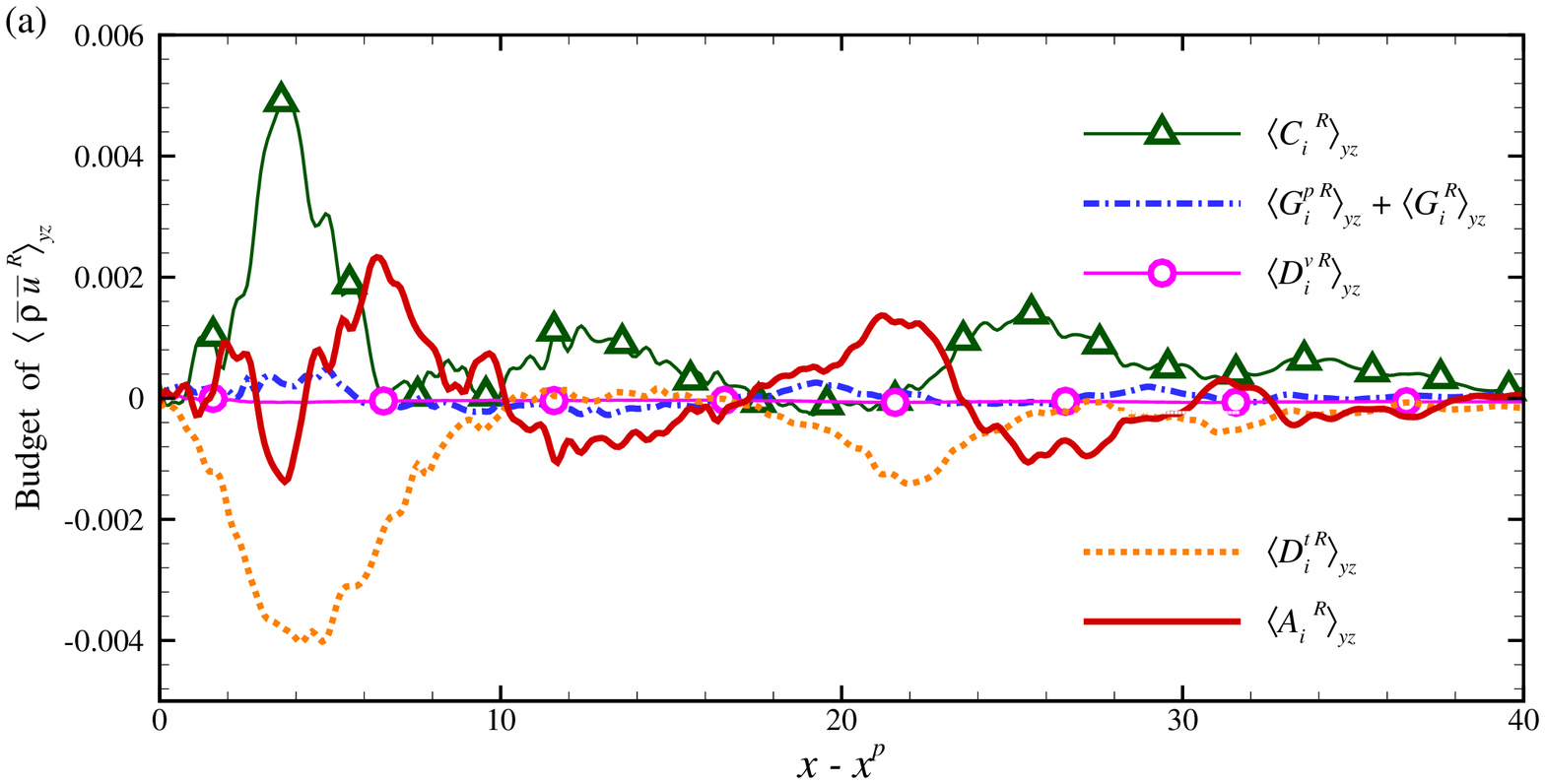}
	\includegraphics[scale=0.6]{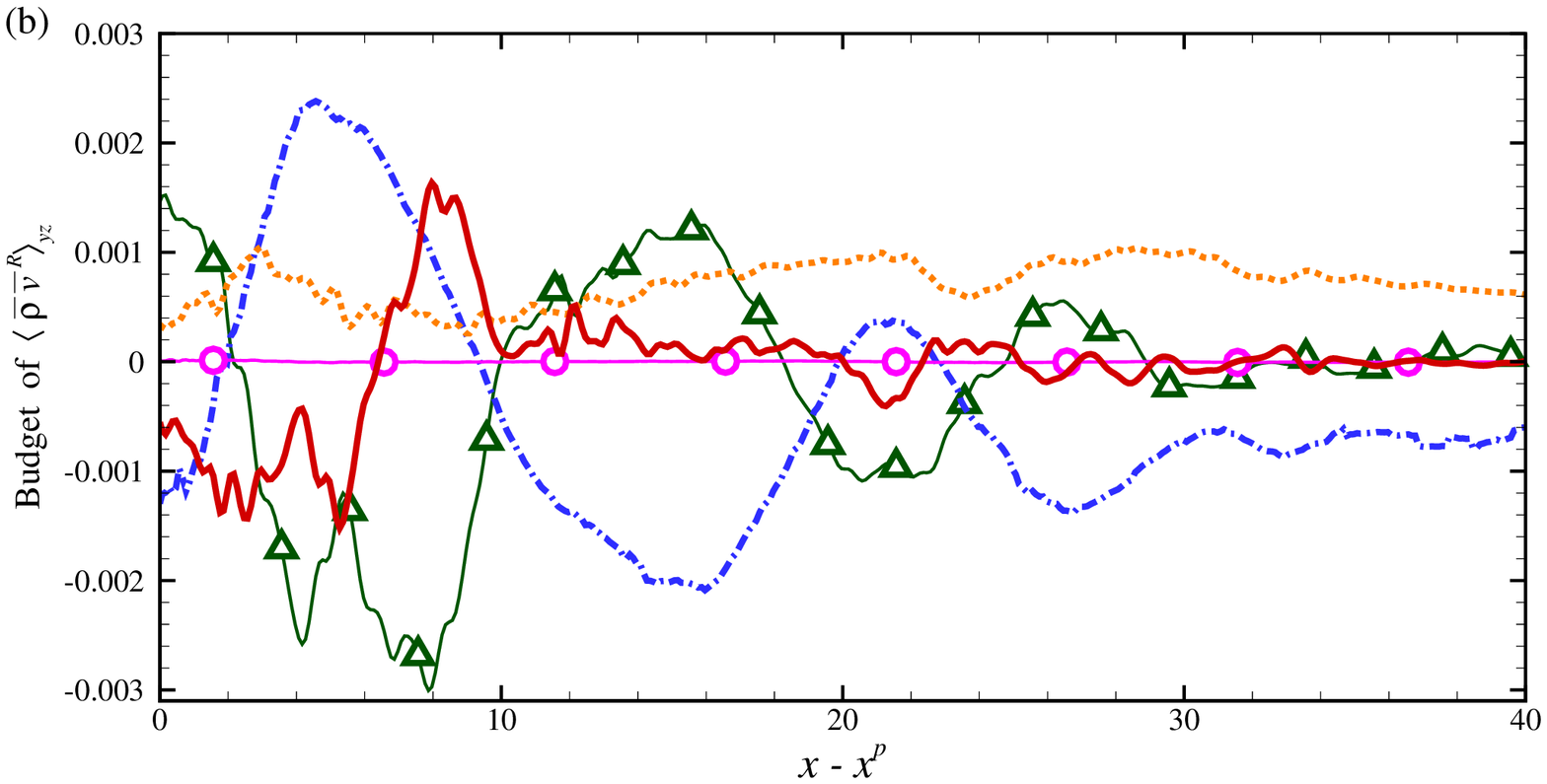}
	\caption{\label{MEAN_terms} 
		Curves of mixed-phase-region-averaged terms in the 
		right hand side of equation (\ref{mean_momentum_equation}) 
		along the streamwise direction. 
		($ \text{Re} = 1600 $ and $ \text{Fr} = 6.4 $)
	}
\end{figure}

Figure~\ref{MEAN_terms} shows the averaged value of each budget term of 
equation (\ref{mean_momentum_equation}) in the mixed-phase region. 
As shown in figure~\ref{MEAN_terms}(a), the convection term $C_1$, 
Reynolds stress term $D_1^t$ and TMF term $A_1$ make dominant contribution 
to the transport of $\overline{\rho} \overline{u}$. 
The convection term $C_1$ and Reynolds stress term $D_1^t$ balance 
with each other near the jet plunging point. 
They decay downstream and the TMF term $A_1$ becomes a dominant term. 
Among the budget terms of $\overline{\rho} \overline{v}$, the summation 
of the gravity term $G_2$ and mean pressure gradient term $G^p$ 
make significant contribution. 
It shows consistency with the variation of the mean vertical velocity
along the streamwise direction. 
The TMF term $A_2$ is important near the jet plunging point 
and decays to a small magnitude for $x - x^p \ge 10.0$. 
The Reynolds stress term $D_2^t$ plays an important role for $x - x^p \ge 10.0$. 
The results shown in figure~\ref{MEAN_terms} indicate that 
the closure of both Reynolds stress and TMF is important 
in the mixed-phase turbulence induced by jet plunging.

\subsection{Turbulent kinetic energy}
\label{subsec_Turbulent_kinetic_energy}

There are different strategies for closing the Reynolds stress 
$ \overline{\rho u_i^{\prime} u_j^{\prime}} $. 
In the single-phase flows, an important strategy is to use the dynamic equation of 
TKE $ k = \frac{1}{2} \overline{\rho u_i^{\prime} u_i^{\prime}} $ for closure, 
such as the $k$--$\varepsilon$ Model~\citep{chienPredictionsChannelBoundaryLayer1982, 
	kaulEffectInflowBoundary2010, 
	kaulEffectInflowBoundary2011} 
and $k$--$\omega$ model~\citep{wilcoxReassessmentScaledeterminingEquation1988,
	menterTwoequationEddyviscosityTurbulence1994,
	spalartEffectiveInflowConditions2007,
	wilcoxFormulationKwTurbulence2008}. 
In the following context, we first analyse the effect of 
the Froude number on the TKE, followed by some discussions 
on if the closure model of TKE in single-phase turbulence 
can be applied to a mixed-phase turbulent flow. 

\begin{figure}
	\centering
	\includegraphics[scale=0.35]{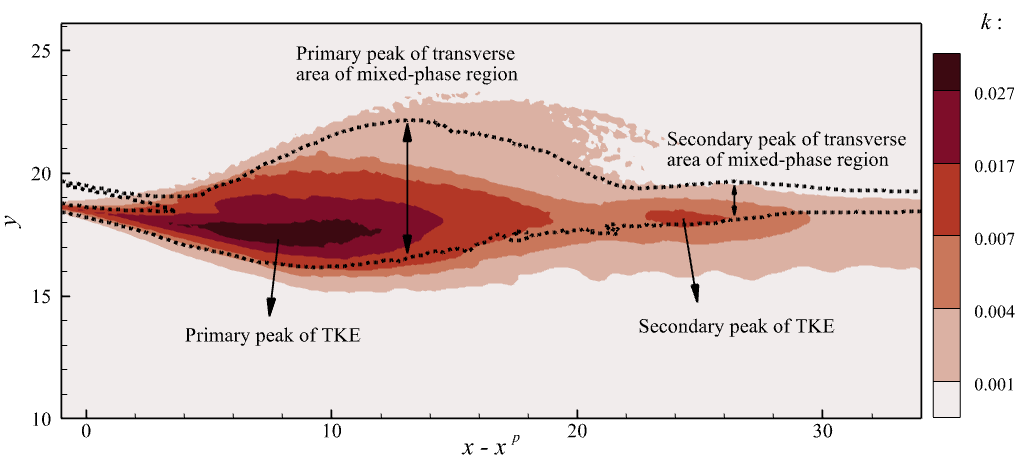}
	\caption{\label{TKE_x-y} 
		Contours of TKE at the mid span. 
		Dotted line represents the edge of the mixed-phase region.
		($ \text{Re} = 1600 $ and $ \text{Fr} = 6.4 $)
	}
\end{figure}

Figure~\ref{TKE_x-y} displays the contours of TKE 
at the mid span for $ \text{Fr} = 6.4 $. 
The figure demonstrates a strong correlation between 
the mixed-phase region and large magnitude of TKE. 
The highest TKE is observed below the mean surface downstream 
near the jet plunging point ($x - x^p = 4.0 \sim 12.0$), 
where the shear between the jet and the pool water is strong. 
At approximately $x - x^p = 24.0$, 
there is a secondary peak of TKE caused by the secondary plunging. 

\begin{figure}
	\centering
	\includegraphics[scale=0.6]{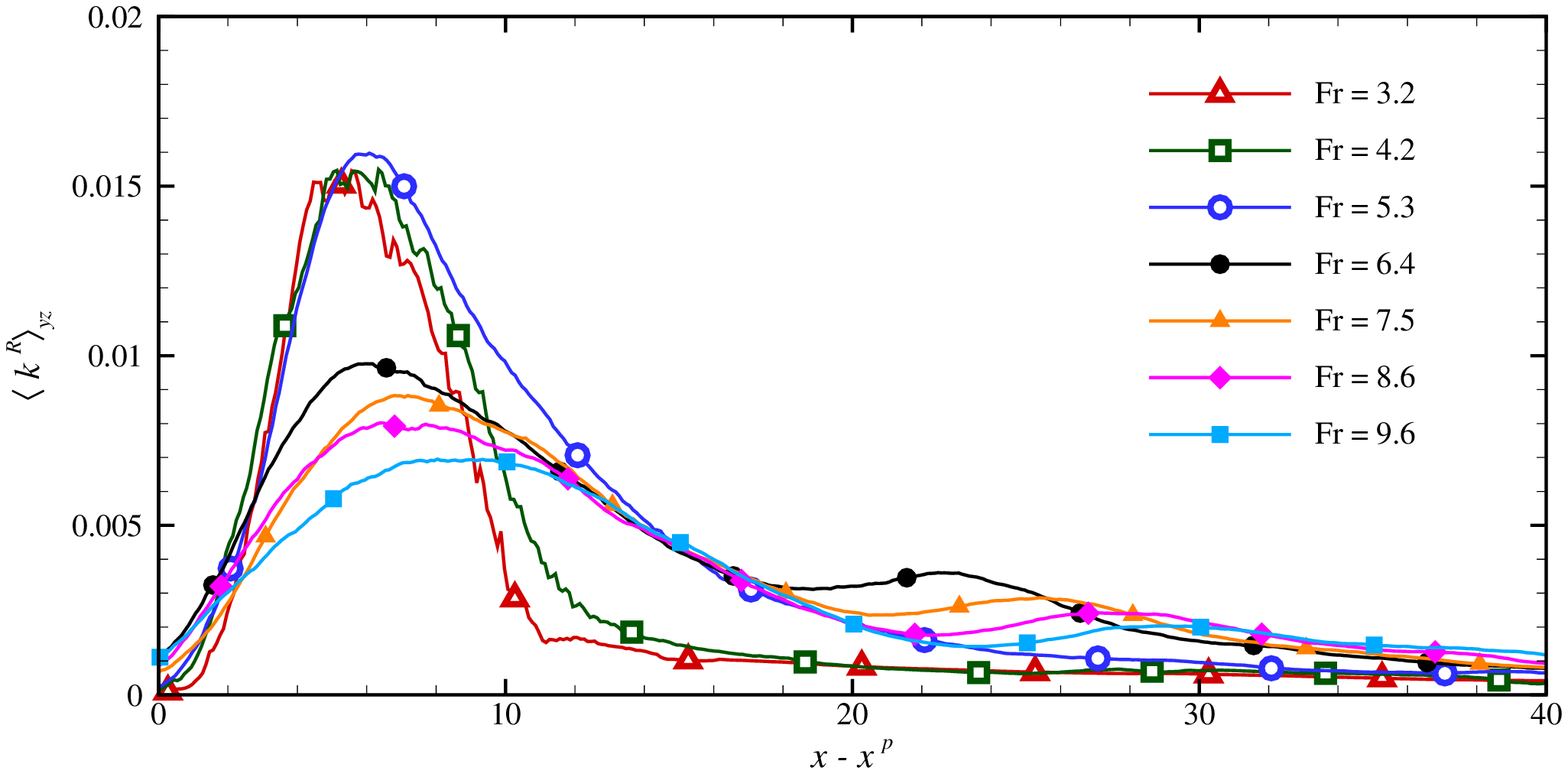}
	\caption{\label{TKE_x-xp_different_Fr} 
		Profiles of average TKE in the mixed-phase region 
		along the streamwise direction in cases of different Froude number.
	}
\end{figure}
\begin{figure}
	\centering
	\includegraphics[scale=0.4]{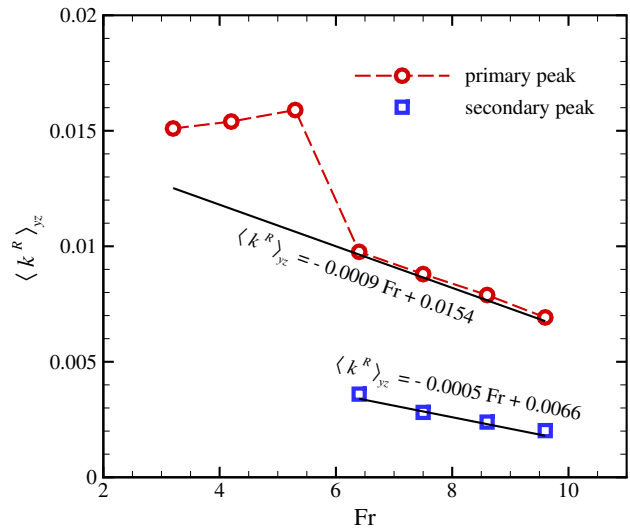}
	\caption{\label{TKE_Fr} 
		Primary peak (red circle symbols) and secondary peak (blue square symbols) variables of 
		the average TKE in varying with Froude number and the fit curve.
	}
\end{figure}

Figure~\ref{TKE_x-xp_different_Fr} compares the streamwise variation of 
the TKE averaged in the mixed-phase region 
$\left\langle k^R \right\rangle _{yz}$ for different Froude numbers. 
At all Froude numbers, a primary peak occurs around $x-x^p = 6.0$. 
For large Froude numbers ($\text{Fr} \ge 6.4$), a secondary peak of TKE occurs. 
Figure~\ref{TKE_Fr} compares the magnitudes of the two peaks at different Froude numbers. 
It is seen that at low Froude numbers, 
the magnitude of the primary peak increases with Froude number. 
As the Froude number increases to $\text{Fr} \ge 6.4$, 
the magnitudes of both primary and secondary peak decrease 
linearly with the Froude number. 
The observations from figure~\ref{TKE_Fr} indicates that the Froude 
number imposes dual effects on TKE. At low Froude numbers, 
the entrained air volume increases and the shear is enhanced near 
the interface as the Froude number increases. 
As a result, the magnitude of TKE increases with Froude number ($\text{Fr} \le 5.3$). 
At larger Froude numbers ($\text{Fr} \ge 6.4$), 
the increase in the Froude number leads to the decrease of 
gravitational potential energy. Consequently, the vertical velocity and 
the plunging angle decrease when the jet hits the free surface. 
Furthermore, at higher Froude numbers, more droplets are generated, 
resulting in secondary plunging. 
In other words, the TKE is distributed in a wider streamwise range 
at higher Froude numbers, resulting in a smaller peak value. 


The transport equation of TKE in variable-density flows is expressed as
~\citep{chassaingVariableDensityFluid2002}

\begin{equation}\label{TKE_transport_equation}
	\frac{\partial k}{\partial t} = 
	C + D^t + P + D^p + D^v + \varepsilon + A. 
\end{equation}

\noindent
Budget terms on the right-hand 
side of equation~(\ref{TKE_transport_equation}) include the convection term $ C $, 
turbulence-diffusion term $D^t$, production term $P$, pressure-diffusion term $D^p$, 
viscous diffusion term $D^v$, dissipation term $\varepsilon$, 
and the TMF-correlation term $A$, defined as 

\begin{align}\label{TKE_terms_def}
	&C = -\frac{\partial\left(k \overline{u}_j\right)}{\partial x_j}, \\
	&D^t = - \frac{\partial\left(\frac{1}{2} \overline{\rho u_i^{\prime} 
			u_i^{\prime} u_j^{\prime}}\right)}{\partial x_j}, \\
	&P = \overline{\rho u_i^{\prime} u_j^{\prime}} 
	\frac{\partial \overline{u}_i}{\partial x_j}, \\
	&D^p = - \frac{\partial \overline{p^{\prime} u_i^{\prime}}}{\partial x_i}, \\
	&D^v = \frac{\partial \overline{\tau_{i j}^{\prime} u_i^{\prime}}}{\partial x_j}, \\
	&\varepsilon = - \overline{\tau_{i j}^{\prime} \frac{\partial u_i^{\prime}}{\partial x_j}}, \\
	&A = \frac{\overline{\rho u_i^{\prime}}}{\text{Fr}^2} \delta_{i2} - \overline{\rho u_i^{\prime}}
	\left(\frac{\partial \overline{u}_i}{\partial t} 
	+ \overline{u}_j \frac{\partial \overline{u}_i}{\partial x_j}\right).
\end{align}

\noindent

\begin{figure}
	\centering
	\includegraphics[scale=0.6]{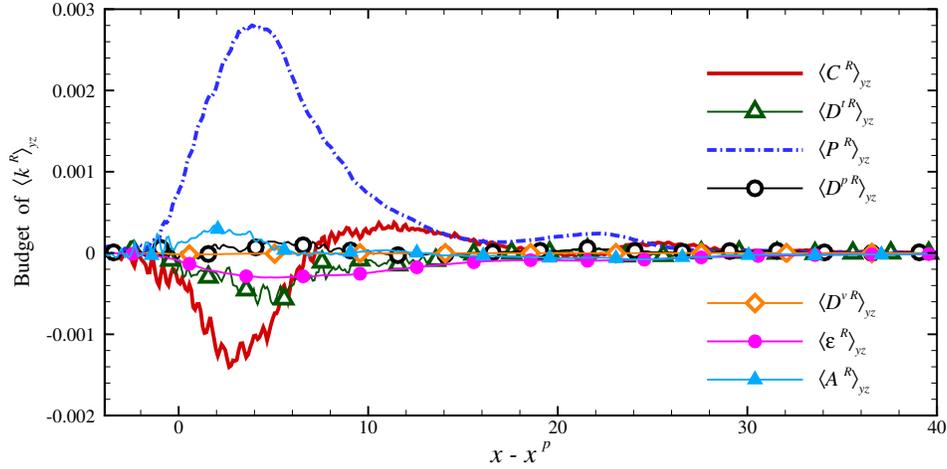}
	\caption{\label{tke_terms_MPR_x} 
		Streamwise variation of the budget terms in transport equation of TKE 
		averaged in the mixed-phase region for 
		$ \text{Re} = 1600 $ and $ \text{Fr} = 6.4 $. 
	}
\end{figure}

Figure~\ref{tke_terms_MPR_x} shows the streamwise variation of the budget 
terms of $\left\langle k^R \right\rangle _{yz}$, TKE averaged in cross-stream 
plane inside the mixed-phase region. 
The results for $ \text{Re} = 1600 $ and $ \text{Fr} = 6.4 $ are shown. 
Similar to many single-phase flows, the balance between the production 
term $P$ and the dissipation term $\varepsilon$ dominates the transport of TKE. 
The convection term $C$ is positive upstream and changes its sign to negative 
downstream, indicating energy convection from upstream to downstream. 
The magnitude of the TMF-correlation term $A$ is smaller than the production term $P$. 
This indicates that the closure model of single-phase turbulence 
can be referenced by the mixed-phase turbulence induced by an plunging jet.

\subsection{Turbulent mass flux and its transports}
\label{subsec_Turbulent_mass_flux_and_its_transports}

Equations~(\ref{mean_momentum_equation})--(\ref{mean_terms_def_A})
show that there are two unclosed terms in the mean momentum equation of 
the mixed-phase turbulence. 
The first term is Reynolds stress tensor and it is usual to consider 
its isotropic part TKE for closure problems. 
Analysis of TKE transport equation shows that its budget remains similar 
to most single-phase turbulent flows.  
The energy generated by the TMF-correlation term is relatively small, 
as such the single-phase flow closure model can be used. 
However, \ref{subsec_Mean_velocity} shows that the TMF term plays 
an important role in the transport of mean momentum. 
Therefore, its closure model is also important for industrial applications. 
\citet{hendricksonWakeThreedimensionalDry2019a} developed an algebraic 
model for TMF based on their iLES data of the wake of 
a three-dimensional dry transom stern. 
Their \textit{a priori} tests showed that it is challenging to obtain 
an ideal correlation between the model and iLES data. 
Another strategy to close the TMF is to develop a dynamic model, 
which requires an analysis of its transport equation. 
In this section, we investigate the Froude number effects on the TMF,  
and the transports of TMF is then analysed. 

\begin{figure}
	\centering
	\includegraphics[scale=0.35]{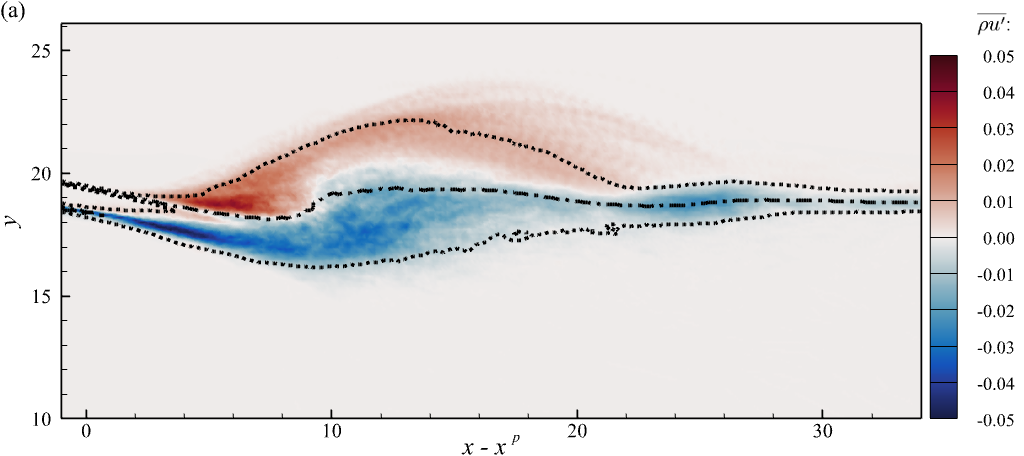}
	\includegraphics[scale=0.35]{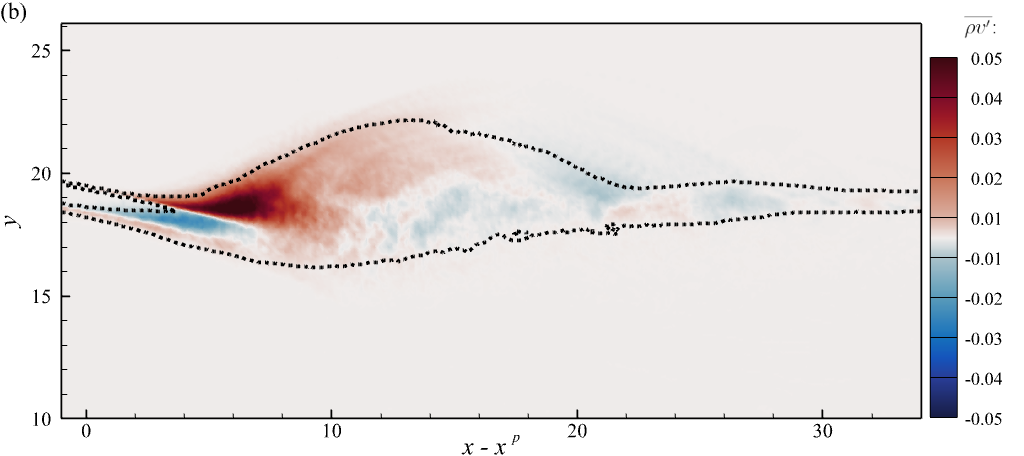}
	\caption{\label{TMF_x-y} 
		Contours of TMF:
		(a) $\overline{\rho u^\prime}$; (b) $\overline{\rho v^\prime}$ 
		at the mid span for $ \text{Re} = 1600 $ and $ \text{Fr} = 6.4 $. 
		The upper and lower dotted lines represent the edge of the mixed-phase region. 
		The dash-dotted line in (a) represents the mean location of free surface. 
	}
\end{figure}

Figures~\ref{TMF_x-y}(a) and (b) show the contours of 
$\overline{\rho u^\prime}$ and $\overline{\rho v^\prime}$, respectively, 
at the mid span for $ \text{Re} = 1600 $ and $ \text{Fr} = 6.4 $. 
The upper and lower dotted lines represent the edge of the mixed-phase region. 
The dash-dotted line in figure~\ref{TMF_x-y}(a) represents the mean 
location of free surface corresponding to $\overline{\psi} = 0.5$. 
As shown in figure~\ref{TMF_x-y}(a), $\overline{\rho u^\prime}$ is positive 
above the mean free surface, indicating the downstream transport of water droplets. 
Below the mean free surface, $\overline{\rho u^\prime}$ is negatively valued, 
corresponding to the downstream transport of bubbles.
From figure~\ref{TMF_x-y}(b), it is observed that near both the 
primary and secondary plunging point, $\overline{\rho v^\prime}$ 
is positive, indicating the air entrainment. Shortly downstream 
the plunging points, $\overline{\rho v^\prime}$ changes its sign 
to negative, indicating the bubbles detrainment.  

\begin{figure}
	\centering
	\includegraphics[scale=0.6]{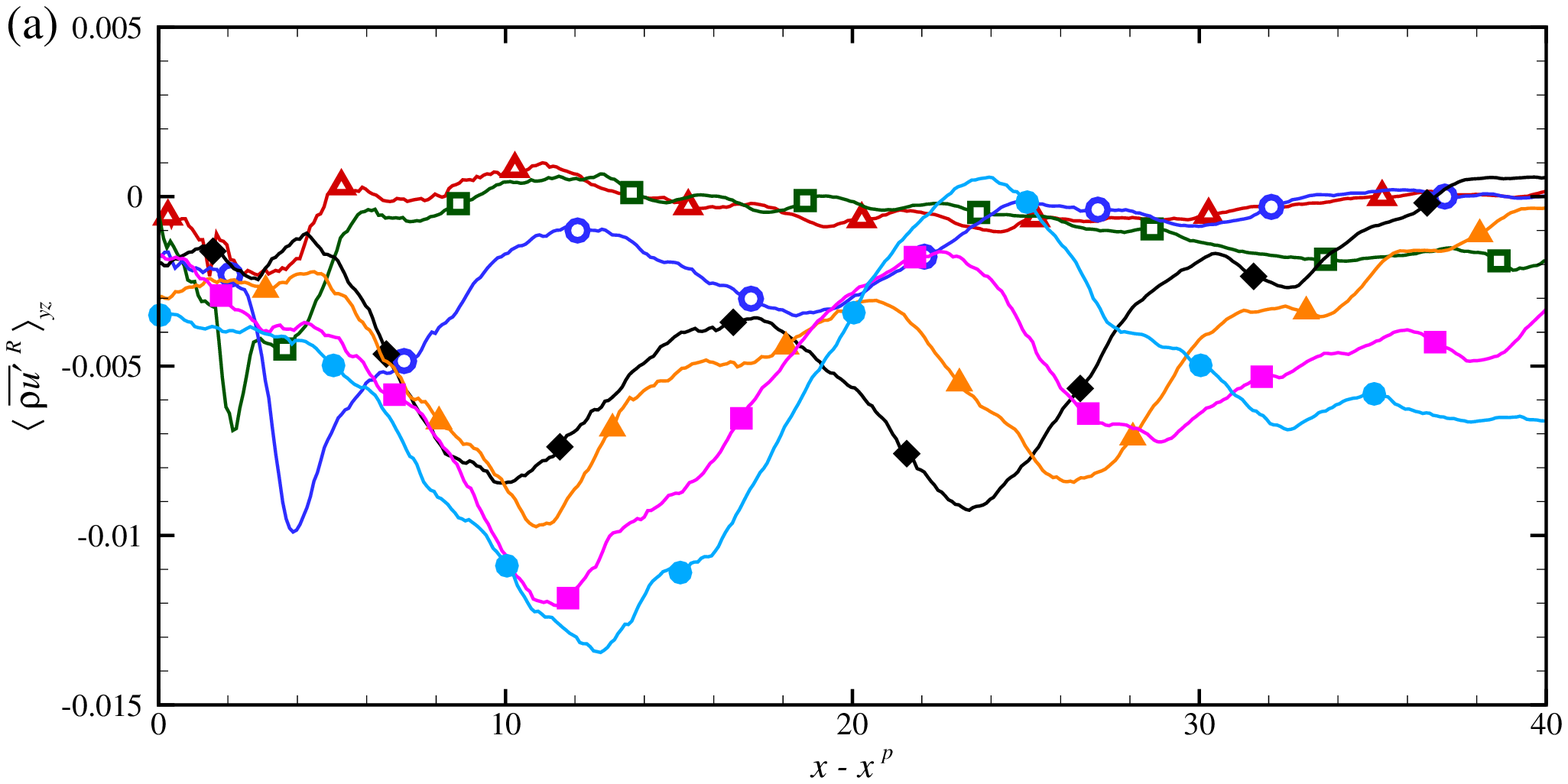}
	\includegraphics[scale=0.6]{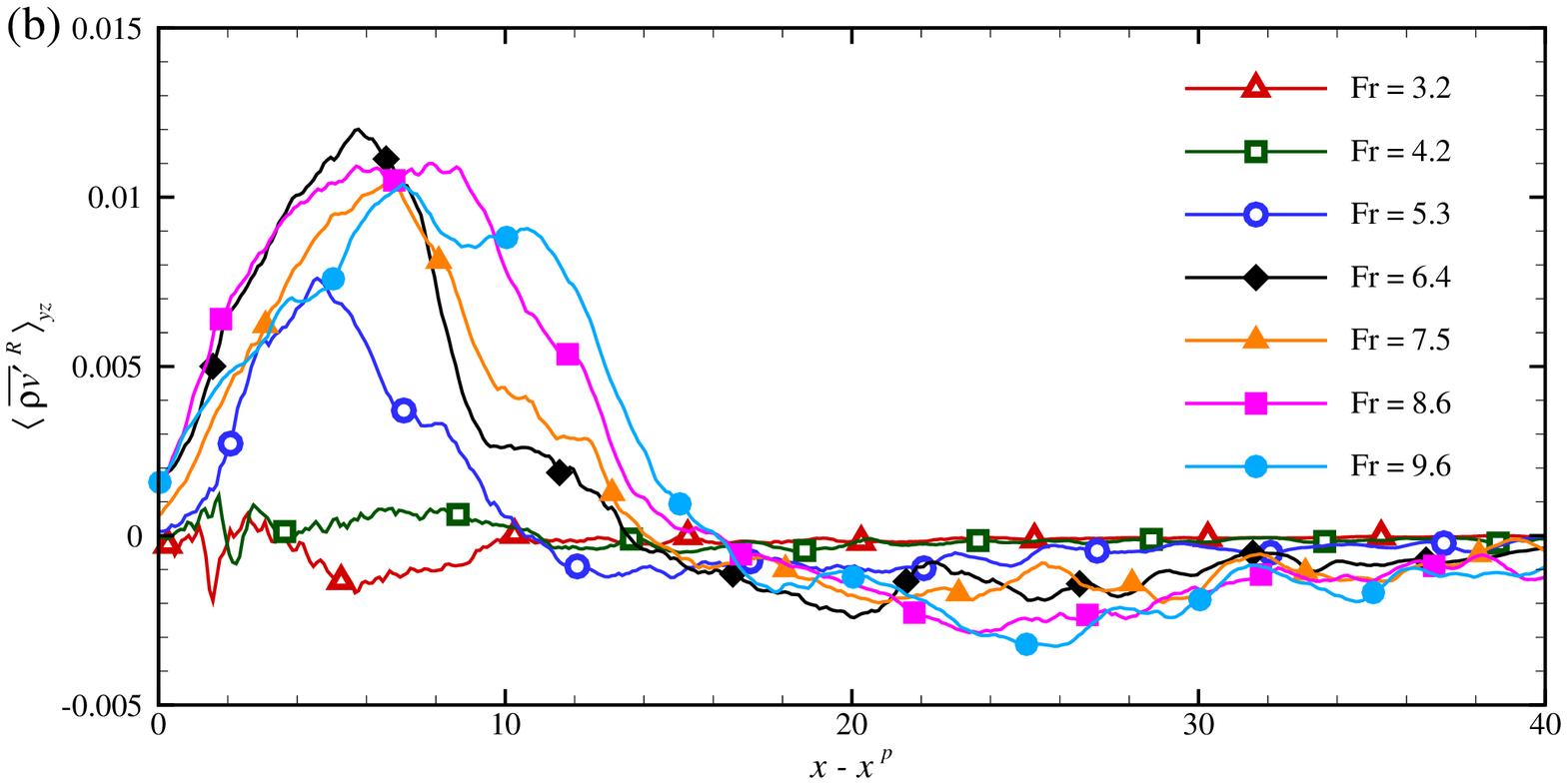}
	\caption{\label{TMF_MPR_x-xp_different_Fr} 
		Profiles of TMF averaged in the mixed-phase region 
		along the streamwise direction in cases of different Froude number:
		(a) $\overline{\rho u^\prime}$, (b) $\overline{\rho v^\prime}$.
	}
\end{figure}

Figure~\ref{TMF_MPR_x-xp_different_Fr} compares the streamwise variation 
of the TMF averaged in the mixed-phase region for different Froude numbers. 
Figure~\ref{TMF_MPR_x-xp_different_Fr}(a) shows that the negative value 
of $\overline{\rho u^\prime}$ dominates in the mixed-phase region. 
Recalling that negative and positive $\overline{\rho u^\prime}$ correspond to 
downstream motion of bubbles and droplets, respectively, 
the negative value of $\left\langle \overline{\rho u^\prime}^R \right\rangle _{yz}$ 
indicates that the downstream transfer of air beneath the free surface is dominant. 
The primary peak of the negatively-valued 
$\left\langle \overline{\rho u^\prime}^R \right\rangle _{yz}$ 
occurs downstream the plunging point. 
Its magnitude increases as the Froude number increases. 
This indicates that more bubbles are convected downstream at higher Froude numbers. 
When $\text{Fr} \ge 6.4$, there exists a secondary peak 
in $\left\langle \overline{\rho u^\prime}^R \right\rangle _{yz}$. 
However, it shows a different trend of the secondary peak as the 
Froude number increases. This is because the convection of droplets 
above the surface balances a part of bubble motion beneath the surface. 
At large Froude numbers, water splash-up induces droplets resulting 
in the decrease in the magnitude of the secondary negatively-valued peak of 
$\left\langle \overline{\rho u^\prime}^R \right\rangle _{yz}$. 

Figure~\ref{TMF_MPR_x-xp_different_Fr}(b) shows that 
$\left\langle \overline{\rho v^\prime}^R \right\rangle _{yz}$ 
is positive near the plunging point, 
indicating air entrainment in this region. 
The magnitude of $\left\langle \overline{\rho v^\prime}^R \right\rangle _{yz}$  
is small at $\text{Fr} = 3.2$ and 4.2. 
It increases as the Froude number increases from 4.2 to 6.4. 
As the Froude number continues to increase, its magnitude decreases slightly, 
but the streamwise range with positive 
$\left\langle \overline{\rho v^\prime}^R \right\rangle _{yz}$ increases. 
This indicates that the air entrainment takes place in a larger streamwise 
region at a higher Froude number. 
A negative peak of $\left\langle \overline{\rho v^\prime}^R \right\rangle _{yz}$ 
occurs for $\text{Fr} \ge 6.4$, caused by the bubble detrainment after 
the secondary plunging. The magnitude of this negatively-valued peak of 
$\left\langle \overline{\rho v^\prime}^R \right\rangle _{yz}$ increases  
as the Froude number increases from 6.4 to 9.6, indicating that more bubbles 
are detrained at higher Froude numbers. 

To investigate the closure of TMF, 
we examine the following transport equation of the TMF: 

\begin{equation}\label{TMF_transport_equation}
	\frac{\partial \overline{\rho u_i^{\prime}}}{\partial t} = 
	C_i + P^{(1)}_i + P^{(2)}_i + D_i + E_i. 
\end{equation}

\noindent
The budget terms on the right-hand side include convection term $C_i$, 
production terms $P^{(1)}_i$ and $P^{(2)}_i$ corresponding to the 
velocity gradient and density gradient, respectively, 
turbulent diffusion $D_i$, and a combining term $E_i$. 
The definitions of these terms are given as 
\begin{align}
	\label{TMF_terms_def_Conv}
	&C_i = -\frac{\partial \left(\overline{\rho u_i^{\prime}} \overline{u}_j \right)}{\partial x_j}, \\
	\label{TMF_terms_def_Prod_1}
	&P^{(1)}_i = -\overline{\rho u_j^{\prime}} \frac{\partial \overline{u}_i}{\partial x_j}, \\
	\label{TMF_terms_def_Prod_2}
	&P^{(2)}_i = -\overline{u_i^{\prime} u_j^{\prime}} \frac{\partial \overline{\rho}}{\partial x_j}, \\
	\label{TMF_terms_def_Diff}
	&D_i = \frac{\partial \overline{\rho} \overline{u_i^{\prime} u_j^{\prime}}}{\partial x_j} 
	- \frac{\partial \overline{\rho u_i^{\prime} u_j^{\prime}}}{\partial x_j}, \\
	\label{TMF_terms_def_Equi}
	&E_i = \left(\overline{\rho} \overline{\frac{1}{\rho} \frac{\partial p}{\partial x_i}} 
	- \frac{\partial \overline{p}}{\partial x_i}\right) 
	+ \left(\frac{\partial \overline{\tau}_{i j}}{\partial x_j} 
	- \overline{\rho} \overline{\frac{1}{\rho} \frac{\partial \tau_{i j}}{\partial x_j}}\right). 
\end{align}

\noindent
It should be noted here that the combining term $E_i$ consists of a 
pressure-gradient part and a viscous-stress part. 
From its expression, it is understood that this term is essentially 
the difference between their time-averaged values 
(i.e. ${\partial \overline{p}}/{\partial x_i}$ and 
${\partial \overline{\tau}_{i j}}/{\partial x_j}$) 
and their density-weighted time-averaged values 
(i.e. $\overline{\rho} \overline{{\rho}^{-1} {\partial p}/{\partial x_i}}$ 
and $\overline{\rho} \overline{{\rho}^{-1} {\partial \tau_{i j}}/{\partial x_j}}$). 
To perform the density-weighted time averaging, 
the instantaneous density needs to be interpolated. 
Owing to the use of sharp-interface treatment in the present simulation, 
the instantaneous density $\rho$ varies sharply across the interface. 
As a result, the interpolation of $\rho$ causes oscillation near the interface, 
resulting in non-physical values. 
Therefore, in the present study, the pressure-gradient and viscous-stress terms 
are combined into one term $E_i$, and its value is determined indirectly as the 
opposite number of the summation of other terms. 
This treatment is supported by the assumption that 
${\partial \overline{\rho u_i^{\prime}}}/{\partial t} = 0$ is attained, 
and sufficiently long time duration is used for performing time averaging. 

\begin{figure}
	\centering
	\includegraphics[scale=0.6]{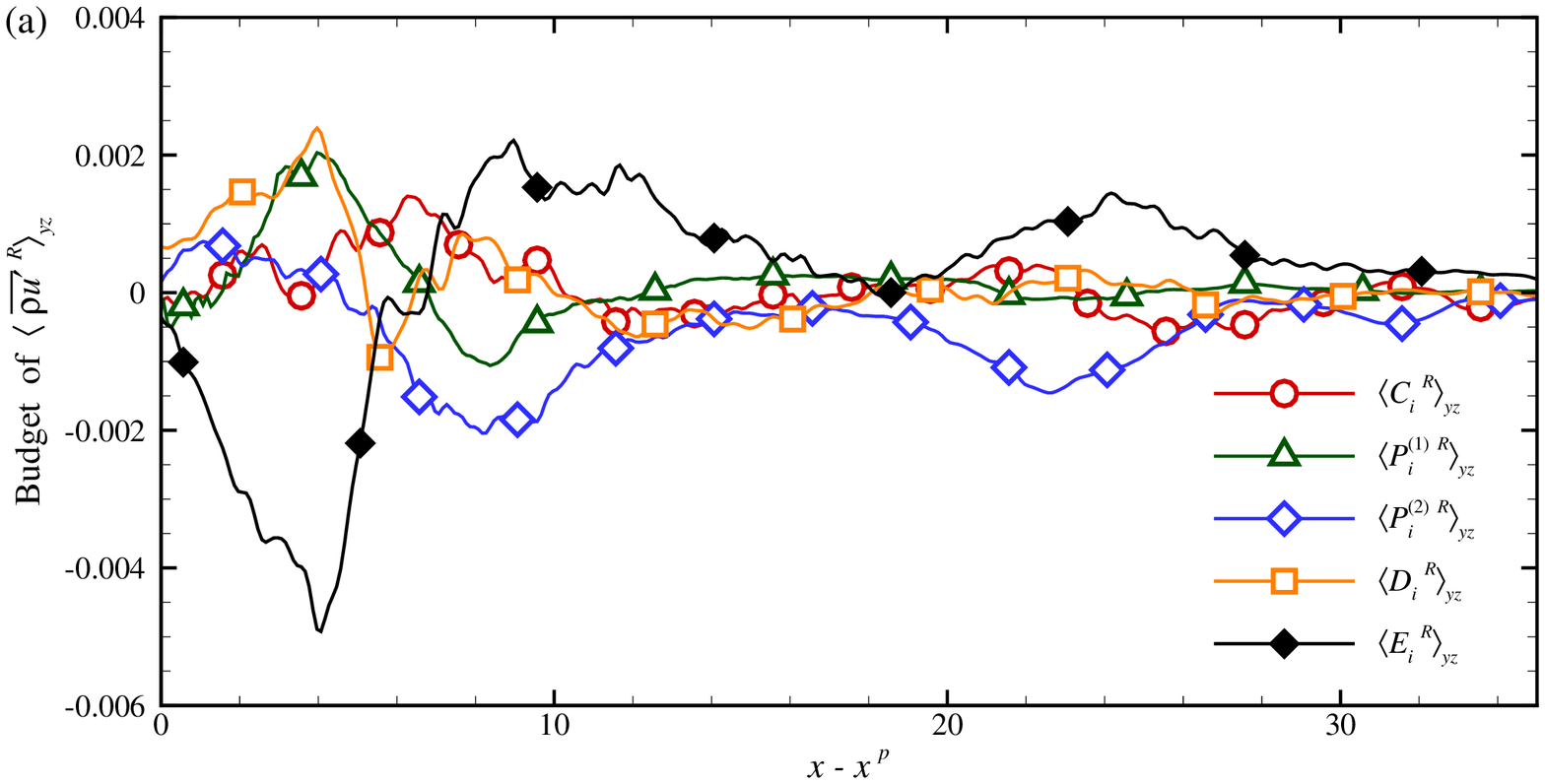}
	\includegraphics[scale=0.6]{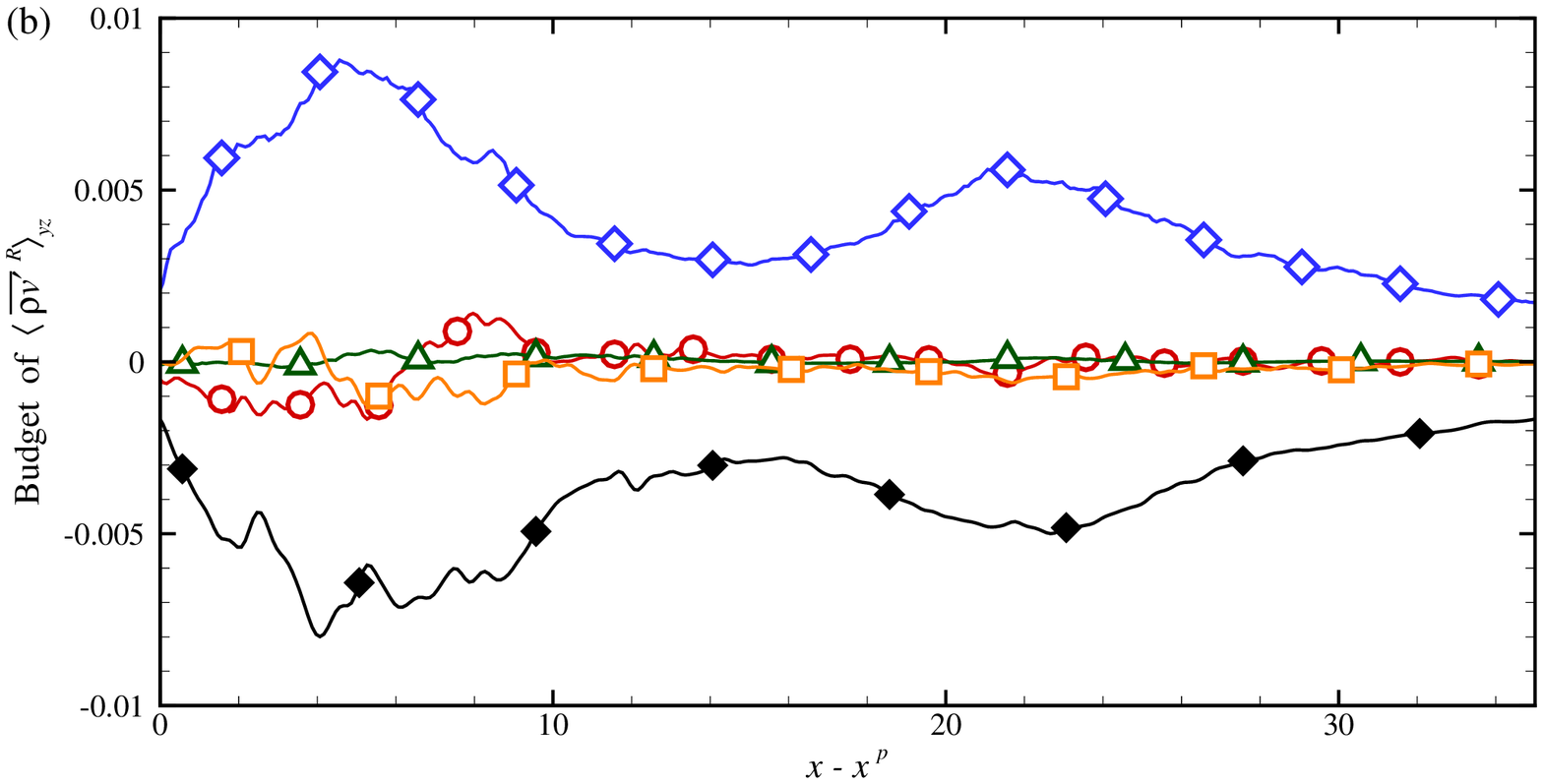}
	\caption{\label{TMF_terms_x-xp} 
		Profiles of average terms on the right-hand side of the TMF
		(a) $\overline{\rho u^\prime}$, (b) $\overline{\rho v^\prime}$ 
		transport equation~(\ref{TMF_transport_equation})  
		in the mixed-phase region along the streamwise direction. 
		($ \text{Re} = 1600 $ and $ \text{Fr} = 6.4 $)
	}
\end{figure}

Figure~\ref{TMF_terms_x-xp} shows the streamwise variation of 
the budget terms of TMF averaged in the mixed-phase region of cross-stream planes. 
From figure~\ref{TMF_terms_x-xp}(a), it is seen that the budget of 
$\left\langle \overline{\rho u^\prime}^R \right\rangle _{yz}$ is influenced 
by all terms near the jet plunging point. 
Downstream, convection term $C_1$, production term $P^{(1)}_1$ 
and turbulent diffusion term $D_1$ decay to a relatively small value, and the
budget is mainly balanced by production term $P^{(2)}_1$ and combining term $E_1$. 
The balance between production term $P^{(2)}_2$ and the combining term $E_2$ also 
dominates the budget of $\left\langle \overline{\rho v^\prime}^R \right\rangle _{yz}$. 

On the right hand side of the transport equation of TMF, the convection 
term $C_i$ and the production term $P^{(1)}_i$ do not require modelling. 
Furthermore, as shown in figure~\ref{TMF_terms_x-xp}, they are not the 
dominant budget terms. Term $D_i$, which can be seen as the diffusion 
effect of velocity fluctuation on TMF, is only important in the transport of 
$\left\langle \overline{\rho u^\prime}^R \right\rangle _{yz}$ near the plunging point, 
and it can be modelled by estimating a characteristic diffusion velocity using TKE. 
Term $E_i$ is an important transportation term, but due to the lack of reliable data 
and deeper understanding, it is currently difficult to establish a closure model for this term. 
Considering the both pressure and viscous stress fluctuations are induced by velocity fluctuations, 
term $E_i$ can be seen as a passive response of the flow field to the change in the other budget terms of the TMF, 
and can be possibly modelled as a diffusion effect induced by an artificial viscous. 
Term $P^{(2)}_i$ is related to the density gradient caused by two-phase mixture, 
which is an important production term and can be seen as the direct source term of TMF. 
Therefore, it is crucial to close term $P^{(2)}_i$ in a dynamic model of TMF. 

%
\begin{figure}
	\centering
	\includegraphics[scale=0.6]{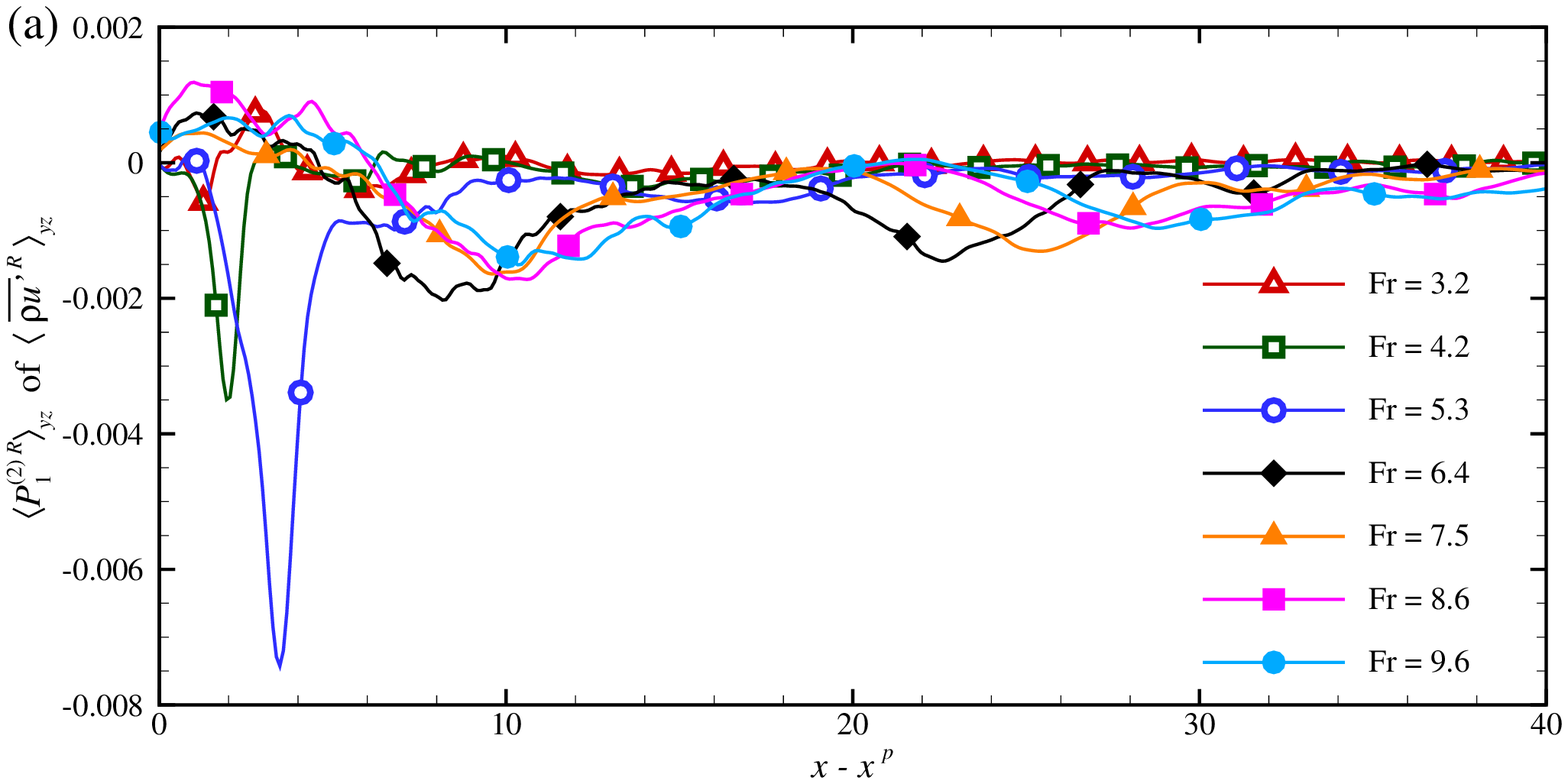}
	\includegraphics[scale=0.6]{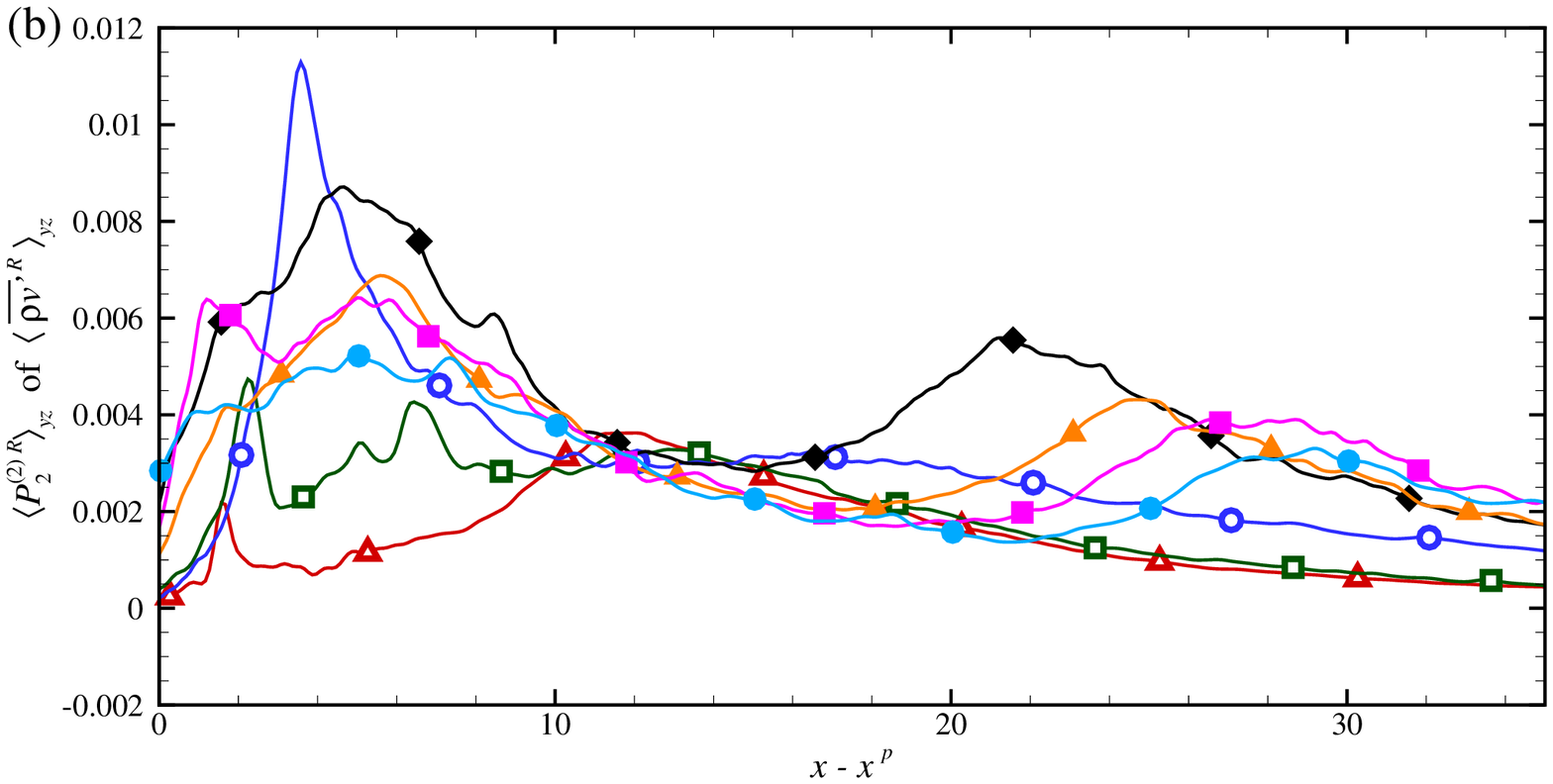}
	\caption{\label{TMF_PROD_DEN} 
		Curves of mean density gradient production term of the TMF 
		(a) $\overline{\rho u^\prime}$, (b) $\overline{\rho v^\prime}$ 
		averaged in the mixed-phase region 
		along the streamwise direction in different Froude numbers. 
	}
\end{figure}

Figure~\ref{TMF_PROD_DEN} compares the streamwise variation of $P^{(2)}_i$ 
averaged in the mixed-phase region in different Froude numbers. 
At lower Froude numbers ($\text{Fr} \le 5.3$), 
$\left\langle P_1^{(2) R} \right\rangle _{yz}$ shows a sharp peak 
near the jet impact point and its absolute value increases with the Froude number. 
At higher Froude numbers ($\text{Fr} \ge 6.4$), 
$\left\langle P_1^{(2) R} \right\rangle _{yz}$ shows two peaks corresponding to 
the two plunging events and they both decrease as the Froude number increases. 
From the comparison between figures~\ref{TMF_PROD_DEN}(a) and \ref{TMF_PROD_DEN}(b), 
it is understood that despite of the opposite sign, 
$\left\langle P_2^{(2) R} \right\rangle _{yz}$ is  similar to 
$\left\langle P_1^{(2) R} \right\rangle _{yz}$ in terms of both 
the variation in the streamwise direction and the Froude number effect. 

The results of the production term of TMF corresponding to the 
density gradient displays a better consistency with TKE than TMF. 
Specifically, there are two peaks of TKE at large Froude numbers 
($\text{Fr} \ge 6.4$) and the magnitudes of both peaks decrease 
as the Froude number increases. Although the TMF also shows two 
peaks at large Froude numbers ($\text{Fr} \ge 6.4$), 
their magnitudes show different trends as Froude number increasing.
Furthermore, the secondary peak value of 
$\left\langle \overline{\rho v^\prime}^R \right\rangle _{yz}$ is negative, 
while both peaks of $\left\langle P_2^{(2) R} \right\rangle _{yz}$ are positive. 
This indicates that $P_i^{(2)}$ can be well modelled by TKE. 

Based on the above analyses, we propose a of $P^{(2)}_i$ as 

\begin{equation}\label{model_equation} 
	-\overline{u_i^{\prime} u_j^{\prime}} \frac{\partial \overline{\rho}}{\partial x_j} = 
	-C_i \overline{u_k^{\prime} u_k^{\prime}} \frac{\partial \overline{\rho}}{\partial x_i}, 
\end{equation}

\noindent
where $C_i$ is the model coefficients. A linear-least squares fit between 
$-\overline{u_i^{\prime} u_j^{\prime}} \frac{\partial \overline{\rho}}{\partial x_j}$ and 
$-\overline{u_k^{\prime} u_k^{\prime}} \frac{\partial \overline{\rho}}{\partial x_i}$ 
in the mixed-phase region is used to determine the model coefficient. 
Model coefficients and conditioned correlation coefficients $R_i$
between the two sides of equation (\ref{model_equation}) at different 
Froude numbers are listed in table~\ref{table_CM_COR1}. 
Coefficients of streamwise and vertical components of equation (\ref{model_equation}) 
are displayed while the spanwise component is not shown because the spanwise 
component of TMF is much smaller than the other two components. 
Correlation coefficient of the vertical component exceeds 0.85 and 
the model coefficient $C_2$ vary little with the Froude number. 
This indicates that $P_2^{(2)}$ can be well estimated by the proposed model. 
Although the accuracy of the streamwise component estimated by the model 
is not as good as the vertical component, correlation coefficients of 
the streamwise component are all above 0.55, which is overall higher than 
the correlation coefficient of the model directly fitting TMF using TKE 
in \cite{hendricksonWakeThreedimensionalDry2019a}. 
%

\begin{table}
	\caption{\label{table_CM_COR1} 
		Model coefficient in equation (\ref{model_equation}) and conditioned correlation coefficient 
		between the two sides of equation (\ref{model_equation}) in the mixed-phase region. 
		Two columns of $ C_i $ and $ R_i $ indicate the streamwise and vertical components 
		of equation (\ref{model_equation}) respectively. 
	}
	\begin{center}
		\begin{tabular}{*{3}{c}*{1}{c|}*{4}{c}*{1}{c|}*{4}{c}}
			\hline
			case & & $ \text{Fr} $ & & & $ C_1 $ & & $R_1$ & & & $ C_2 $ & & $ R_2 $ \\
			\hline
			3 & & 3.2 & & & 0.49 & & 0.67 & & & 0.30 & & 0.91 \\
			4 & & 4.2 & & & 0.60 & & 0.56 & & & 0.33 & & 0.87 \\
			5 & & 5.3 & & & 0.58 & & 0.79 & & & 0.35 & & 0.96 \\
			6 & & 6.4 & & & 0.32 & & 0.56 & & & 0.35 & & 0.92 \\
			7 & & 7.5 & & & 0.39 & & 0.69 & & & 0.33 & & 0.93 \\
			8 & & 8.6 & & & 0.46 & & 0.75 & & & 0.33 & & 0.95 \\
			9 & & 9.6 & & & 0.45 & & 0.75 & & & 0.32 & & 0.94 \\
			\hline
		\end{tabular}
	\end{center}
\end{table}

\section{Conclusion}
\label{sec_Conclusion}

In the present study, we preformed high-resolution interface-resolved LES to study the mixed-phase turbulence induced by a water jet plunging into a quiescent pool. 
In total nine cases were conducted.
The Reynolds number ranged from $1.6 \times 10^3$ to $1.6 \times 10^5$ and the Froude number ranged from $3.2$ to $9.6$ in these cases. 
By comparing the results of different cases, it was discovered that the effect of the Reynolds number on turbulent statistics was less significant than the Froude number.
As a result, this paper mainly focused on the effects of the Froude number on turbulent statistics.
The simulation results showed that increasing the Froude number led to the increase in the  area of the mixed-phase region. 	
To facilitate comparison among results of different cases, a conditioned average over the cross-sectional area inside the mixed-phase region was adopted.

The mean velocity averaged in the mixed-phase region varies non-monotonically with the Froude number. 
As the Froude number increases for 3.2 to 9.6, the magnitude of the mean streamwise velocity reaches its maximum at $Fr=6.4$. 
The mean vertical velocity shows a single negatively-valued peak for low Froude numbers.  
As the Froude number increases to $Fr\ge6.4$, water splash-up and secondary plunging take place, causing respectively a positively valued peak and a secondary negatively valued peak in the mean vertical velocity along the streamwise direction. 
The complex behaviour of the mean velocity is correlated to the nonlinear effects corresponding to the turbulent fluctuation.
In mixed-phase turbulence, there exist two unclosed terms in the Reynolds-averaged mean momentum equation, called the 
Reynolds stress $\overline{\rho u_i^{\prime} u_j^{\prime}}$ 
and turbulent mass flux $\overline{\rho u_i^{\prime}}$. 
The analysis of the mean momentum equation showed that both the Reynolds stress and turbulent mass flux are important terms that require closure models. 

To discuss the closure problem of the Reynolds-averaged mean momentum equation, we analysed the TKE, TMF
and their transport equations. 
Our simulations showed that the TKE also varies non-monotonically with an increasing Froude number. 
At low Froude numbers, the TKE shows a singly peak near the plunging point.
The magnitude of this TKE peak increases as the Froude number increases from 3.2 to 5.3.
As the Froude number increases to $Fr\ge6.4$, the secondary plunging induces a secondary peak in the TKE. 
The magnitudes of both primary and secondary peaks of the TKE decreases as the Froude increases from 6.4 to 9.6. 
The analysis of the transport equation of TKE showed that it is dominated by the balance between production and dissipation. 
The convection and turbulent diffusion is mainly responsible for the spatial transport of TKE. 
Transports characteristics of TKE indicate that its closure model for single-phase turbulence can be used in the mixed-phase turbulence induced by the plunging jet. 

The TMF term is an additional unclosed term in the Reynolds-averaged mean momentum equation of mixed-phase turbulence. 
Its value is zero in the single-phase region, while in the mixed-phased region. 
The streamwise component of TMF is positive above the mean water elevation and is negative below the mean water elevation,  corresponding to the streamwise convection of droplets in the air and bubbles in the water, respectively. 
When the streamwise component of TMF is averaged in the mixed-phase region, its value is negative, indicating that the convection of bubbles dominates the TMF in the streamwise direction.
As the Froude number increases, the magnitude of the streamwise component of TMF increases, corresponding to enhanced downstream convection of bubbles. 
Positive and negative vertical component of TMF occurs alternatively along the streamwise direction. 
Owing to air entrainment, the vertical component of TMF is positive near the plunging points.  
In the downstream, the bubbles are detained, causing negative vertical component of TMF. 
As the Froude number increases, the air entrainment is enhanced.  
This is characterized by the expansion of the streamwise region with positive vertical TMF. 
Meanwhile, the air detrainment in the downstream is also enhanced as reflected by the increase in the negative peak of the vertical TMF.
In a further analysis of the transport equation of TMF, it is discovered that the production term corresponding to the density gradient shows consistency with the TKE. 
Based on this finding, a model of this production term is proposed.
The \textit{a priori} test shows satisfactory correlation between the modeled value and the LES data.

To close this paper, we compare the main findings of the present study 
on the mixed-phase turbulence generated by plunging jet with the results of
~\cite{hendricksonWakeThreedimensionalDry2019a} on the mixed-phase turbulence 
in a wake flow. The purpose of providing this comparison is to evaluate if the main conclusions of the present study are potentially common 
for different mixed-phase turbulence, or they are 
special for the plunging jet. 
In the present study, it is discovered that the transport of TKE is dominated by the balance between production and 
dissipation. 
This is similar to the wake flow. 
However, the TKE generated by TMF shows less significance in the plunging jet than in the wake flow. 
As pointed out by \cite{hendricksonWakeThreedimensionalDry2019a}, the model of TMF is needed to close the TKE transport equation. 
In terms of the closure model of the TMF, the results of both present study and ~\cite{hendricksonWakeThreedimensionalDry2019a} showed that the correlation between TMF and TKE is not strong.
Based on this point, we further investigated the transport equation of TMF. 
We observed that the production term of TMF is in good agreement with TKE. 
This is a new finding of the present study, leading to a closure model of the production term of TMF, which is potentially useful for future development of a dynamic model of TMF. 


\bibliographystyle{model1-num-names}
\bibliography{mybibfile}

\end{document}